\documentclass[%
reprint,
nofootinbib,
amsmath,amssymb,
aps,
amsart
floatfix,
]{revtex4-1}

\usepackage{graphicx}
\usepackage{dcolumn}
\usepackage{bm}
\usepackage{amsmath ,amssymb,mathrsfs,wasysym,amsthm, textcomp, mathtools,graphicx,stmaryrd,lipsum,amscd}
\DeclareMathAlphabet{\mathpzc}{OT1}{pzc}{m}{it}

\usepackage{hyperref}

\hypersetup{%
	colorlinks=false,
	pdfborderstyle={/S/U/W 2}
}

\newcommand{\Rea}{\mathbb{R}}
\newcommand{\Comp}{\mathbb{C}}
\newcommand{\Id}{\hat{\mathbb{I}}}
\newcommand{\Nat}{\mathbb{N}}

\newcommand{\Hi}{\mathcal{H}}

\newcommand{\st}{\mspace{5mu} | \mspace{5mu} }

\newcommand{\boundo}{\mathcal{B}(\mathcal{H})}

\newcommand{\compo}{\mathcal{B}_{\infty}(\mathcal{H})}
\newcommand{\hso}{\mathcal{B}_{2}(\mathcal{H})}
\newcommand{\tco}{\mathcal{B}_{1}(\mathcal{H})}

\newcommand{\borel}{\mathscr{B}(\mathbb{R})}
\newcommand{\borell}[1]{\mathscr{B}(\mathbb{R}^{#1})}
\newcommand{\borelx}{\mathscr{B}(\mathsf{X})}

\newcommand{\E}{\mathcal{E}}

\newcommand{\Ex}{\mathbb{E}}

\newcommand{\X}{\mathsf{X}}

\newcommand{\PS}{(\Omega,\E,P)}

\newcommand{\sss}{\mathbf{S}}
\newcommand{\Int}{\mathbb{Z}}
\newcommand{\Spb}{\mathbb{S}^B}

\newcommand{\M}{\mathsf{M}}
\newcommand{\Ltco}{\mathcal{B}_1^{loc}(L_2(\Rea^d))}
\newcommand{\Spc}{\mathbb{S}^C}

\begin{document}
	
	\newtheorem{definition}{Definition}
	\newtheorem{proposition}{Proposition}
	\newtheorem{theorem}{Theorem}
	\newtheorem{lemma}{Lemma}
	\newtheorem{corollary}{Corollary}
	\newtheorem{assumptions}{Assumptions}
	\newtheorem{assumption}{Assumption}
	
	\renewcommand{\thefootnote}{\fnsymbol{footnote}}
	
	\preprint{APS/123-QED}
	
	\title{On non-commutativity in quantum theory (III): \\determinantal point processes and non-relativistic quantum mechanics.}
	
	\author{Curcuraci Luca}
	\affiliation{Department of Physics, University of Trieste, Strada Costiera 11 34151, Trieste, Italy \\ Istituto Nazionale di Fisica Nucleare, Trieste Section, Via Valerio 2 34127, Trieste, Italy}
	\email{ Curcuraci.article@protonmail.com; \\
		luca.curcuraci@phd.units.it }

	\date{\today}
	
	\begin{abstract}
		
		
		This article concludes our critical analysis on the role of non-commutativity in quantum theory. After a brief introduction of the necessary notions on point processes, we re-analyse model B proposed in \emph{On non-commutativity in quantum theory (II): toy models for non-commutative kinematics}, using the point process theory. This viewpoint allows to generalize and modify the space process of model B in a simple manner, obtaining a new model (model C).  This new model allows the recovery of non-relativistic quantum mechanics in a suitable limit.
	\end{abstract}
	
	\maketitle
	
	\tableofcontents
	
	\section{Introduction}\label{sec0}
	
	In \cite{LC2}, we started to develop a series of models with the scope of deriving the commutation relations between position and momentum of non-relativistic quantum mechanics. The most promising attempt is model B. There, space is represented by a random distribution of points on $\Rea$ each of which evolves in times as a Wiener process. The position of the particle is simply the coordinate of the point of space where the particle is, while the velocity is the ratio between the space covered by the particle in a given time interval, and the duration of such time interval. We showed that when we remove space from the model the probability theory describing the particle is non-commutative. In particular, the operators associated to the position and the velocity of the particle, over a common Hilbert space do not commute. Despite this nice feature, a direct comparison with non-relativistic quantum mechanics cannot be done. The essential problem in model B is that the Hilbert space obtained is \emph{not separable in general}, while in ordinary non-relativistic quantum mechanics the Hilbert space is always separable. The goal of this paper is to show that we can suitably modify model B in order to solve this separability problem. This will be done by exploiting our freedom in the choice of the space process used in the model. A comparison with non-relativistic quantum mechanics will then be possible and, in particular, we will be able to derive the commutation relations between position and momentum operators. 
	
	The article is organised as follows. In section \ref{sec1} we review some basic concepts about point processes, which enable to give a better description of the space process. We also introduce a particular kind of point process: the determinantal point process.  We will conclude this section briefly describing a diffusion process which can be constructed from a determinantal point process. In section \ref{sec2} we re-analyze model B  using the notions and structures introduced in section \ref{sec1}. This enables us to generalize such a model from 1-D to 3-D in a straightforward manner. Finally, in section \ref{sec3}, thanks to the point process structure found, a new model is proposed, model C, where we will use as space process a determinantal point process. This model allows a direct comparison with quantum mechanics. In particular, it is capable to reproduce the non-commutativity between the position and momentum operators of ordinary quantum mechanics. We conclude this work with a general review of the models proposed, a list of possible critical points and possible directions for future research.

	\section{Point processes}\label{sec1}
	
	The goal of this section is to make the reader familiar with the notion of point process. We will introduce the general notions and basic objects used to describe a point processes. Then we focus our attention on a particular class of point process which exhibit very interesting features for the solution of the separability problem.
	
	\subsection{General structure}\label{sec1.1}
	
	A simple introduction to the general theory of point processes can be found in \cite{baddeley2007spatial}, while a more systematic approach is presented in \cite{daley2007introduction}. These two are the main references for the concepts introduced here.
	
	Before starting with the formal mathematical description, let us first explain what point processes are and how they are described. Let $\X$ be a $d$-dimensional space and let $\xi$ label a point process on it. Loosely speaking, $\xi$ can be thought as a collection points of $\X$ randomly chosen according to some probability distribution. To describe it, on a $d$-dimensional space, the most natural way is by counting the number of points that fall in a given subset of $\X$. Let $B \subset \X$ and let $N_{\xi}(B)$ be the number of points of $\xi$ inside $B$. It is not difficult to imagine that we can completely describe the point process $\xi$ by knowing $N_{\xi}(B)$ for any subset $B$ of $\X$, i.e. by knowing the collection $\{N_{\xi}(B)\}_{B \subset \X}$. More precisely, we can say that $\{N_{\xi}(B)\}_{B \subset \X}$ contains enough information to recover the position of all points of $\xi$. Note that this is true only if the points of $\xi$ are not too dense: if $N_{\xi}(B) = \infty$ for any $B \subset \X$ we are not able to reconstruct the single points locations by a counting technique. Let us now formalise this idea.
	\begin{definition}
		Let $\X$ be a locally-compact second countable Hausdorff space and let $\borelx$ be the Borel $\sigma$-algebra on it. Let $N: \borelx \rightarrow \Nat$ be a non-negative integer valued counting measure on $\X$. When $N(B) < \infty$ for any $B \in \borelx$ bounded, we say that the counting measure is \emph{boundely finite}. The set
		\begin{equation*}
		\begin{split}
		\mathcal{N}_{\X} := \{ N(\cdot) | N(B)< \infty &\mbox{ for all bounded } B \in \borelx\},
		\end{split}
		\end{equation*}
		is called \emph{space of all the boundely finite counting measures on $\X$}.
	\end{definition}
	The space $\mathcal{N}_{\X}$ can be equipped with the vague topology, which allows to define open sets that can be used to construct a Borel $\sigma$-algebra $\mathscr{B}(\mathcal{N}_{\X})$.
	\begin{definition}\label{PP-generaldef}
		Let $(\Omega,\E,P)$ be a probability space, $\X$ be a locally-compact second countable Hausdorff space and $\mathcal{N}_{\X}$ be the space of all the boundely finite counting measure on it. The $P$-measurable map $\xi: \PS \rightarrow (\mathcal{N}_{\X}, \mathscr{B}(\mathcal{N}_{\X}))$ defined as
		\begin{equation*}
		\xi: \omega \mapsto \{N_{\xi}(B; \omega)\}_{B \in \borelx},
		\end{equation*}
		is called \emph{point process} over $\X$.
	\end{definition}
	Let us explain better how this definition fits with the idea explained in the beginning. Given $\omega \in \Omega$ the realisation $\xi(\omega)$ of the point process is fixed. Then the number of points of $\xi(\omega)$ that fall in $B$ is $N_{\xi(\omega)}(B) = N_{\xi}(B;\omega)$. Since we have this information for any Borel set $B \subset \X$ we can reconstruct the whole collection of points of $\xi(\omega)$. If no confusion arises, we will in general omit the dependence of $\xi$ on $\omega$. Note that in general, such a collection of points cannot be regarded as a discrete random subset of $\X$. Indeed, there can be point processes whose points may overlap, namely $N_{\xi}(\{x\}) = m > 1$ for some $x \in \X$, and others where this does not happens.
	\begin{definition}\label{definition-simple}
		A point process $\xi$ on $\X$ is said \emph{simple} if $N_{\xi}(\{x\}) \leqslant 1$ for all $x \in \X$.
	\end{definition} 
	In case of simple point processes, $\xi$ can be represented as the (random) subset $\xi:=\{x_1, x_2, \cdots\} \subset \X$ without loosing any information. Such collection of points is also called \emph{configuration} of $\xi$. In general we may always represent the point process $\xi$ as a collection of points, i.e. $\xi:= \{x_1,x_2, \cdots\}$, but when the process is not simple they do not form a subset of $\X$ because there can be $x_i = x_j$ for some $i, j$: this information is lost if $\xi$ is thought as a set.
	\begin{definition}
		Let $\xi$ be a point process on $\X$. If $N_{\xi}(\X) = M \leqslant \infty$ then $\xi$ is said \emph{finite}, while if $N_\xi(\X) = \infty$ the point process is said \emph{locally-finite}.
	\end{definition}
	Note that any point process which is locally finite on $\X$, induces a finte point process on any compact subset $\Lambda \subset \X$ by construction.
	
	Let us now explain how a point process is described from the statistical point of view.    Consider a \emph{finite} point process $\xi$ on $\X$, and the probability distibution $\mu_{\xi} := P \circ \xi^{-1}$ induced on $(\mathcal{N}_{\X}, \mathscr{B}(\mathcal{N}_{\X}))$. When the distribution $\mu_{\xi}$ is used to compute the probabilities of events like
	\begin{equation}\label{JM-event}
	\{N(A_1) = n_1\} \cap \cdots \cap\{N(A_r) = n_r\},
	\end{equation}
	with $A_i \in \borelx$ and $n_i \in \Nat$ for all $i =1, \cdots, r$, $\mu_{\xi}$ takes a special name.
	\begin{definition}
		Let $\xi: \PS \rightarrow (\mathcal{N}_{\X}, \mathscr{B}(\mathcal{N}_{\X}))$ be a point process, the probability distributions
		\begin{equation*}
		\begin{split}
		\mu_{\xi} (A_1, n_1; \cdots;& A_r, n_r) :=\\
		P[N_\xi(& A_1) = n_1, \cdots, N_\xi(A_r) = n_r],
		\end{split}
		\end{equation*}
		where $A_i \in \borelx$ and $n_i \in \Nat$ for all $i =1, \cdots, r$, are called \emph{finite-dimensional distributions} (or \emph{fidi}s) of the point processes $\xi$.
	\end{definition}
	The importance of the fidis for the description of a point process is encoded in the following theorem\cite{baddeley2007spatial}.
	\begin{theorem}
		Let $\xi :(\Omega,\E,P) \rightarrow (\mathcal{N}_{\X}, \mathscr{B}(\mathcal{N}_{\X}))$ and $\eta: \PS \rightarrow (\mathcal{N}_{\X}, \mathscr{B}(\mathcal{N}_{\X}))$ be two point process on $\X$. If all the fidis of $\xi$ and $\eta$ coincide, then $\xi$ and $\eta$ have the same distribution.
	\end{theorem}
	This means that, if the fidis of two point processes are the same then they are equal in distribution, namely $\xi \overset{d}{=} \eta$. Typically point processes are not described in terms of fidis directly, but by using two quantities, the Janossy measure and the moment measure, from which the fidis can be derived. Let us introduce the  Janossy measure. Consider the event \eqref{JM-event}, used to compute the fidi $\mu_{\xi} (A_1, n_1; \cdots; A_r, n_r) $ and assume that $\{A_i\}_{i = 1}^r$ is a \emph{finite partition of $\X$} (i.e. $\cup_{i =1}^r A_i = \X$ and $A_i \cap A_j = \{\varnothing\}$ for  any $i \neq j$). When this event happens, the point process contains exactly $n = n_1 + \cdots + n_r$ points. This observation allows us to write the following:
	\begin{equation*}
	\mu_{\xi}(A) = \sum_{n = 0}^{\infty} p_n \Pi_n (A)
	\end{equation*}
	where $A \in \mathscr{B}(\mathcal{N}_{\X})$ is a generic event and 
	\begin{enumerate}
		\item[i)] $\{p_n\}_{n = 0}^\infty$ are the probabilities that the point process has exactly $n$ points, thus they fulfil the normalisation condition
		\begin{equation}\label{JM-normcond}
		\sum_{i = 0}^\infty p_n = 1;
		\end{equation}
		\item[ii)] $\Pi_n(A)$ a probability distribution on $\X^{(n)}$, i.e. the $n$-fold product space $\X \times \cdots \times \X$, which can be interpreted as the probability distribution of the position of the points of $\xi$, given that their number is exactly $n$.
	\end{enumerate}
	To implement indistinguishability of the points of $\xi$, the joint probability distribution $\Pi_n$ should assign equal weight to any permutation of the coordinates $(x_1,\cdots,x_n)$. If $\Pi_n$ is not so, we can always implement indistinguishability by introducing the symmetrised form
	\begin{equation*}
	\Pi^{sym}_n (A_1 \times \cdots \times A_n) = \frac{1}{n!} \sum_{\sigma \in P_n} \Pi_n(A_{\sigma(1)} \times \cdots \times A_{\sigma(n)}) 
	\end{equation*}
	where $P_n$ is the set of all the permutations of $n$ elements and $A_1, \cdots, A_r \in \borelx$ form a finite partition of $\X$.
	\begin{definition}
		Let $\xi: (\Omega,\E,P) \rightarrow (\mathcal{N}_{\X}, \mathscr{B}(\mathcal{N}_{\X}))$ be a point process on $\X$. Taken a finite partition $A_1, \cdots, A_n \in \borelx$ the ($n$-th) \emph{Janossy measure} of $\xi$ is defined as
		\begin{equation*}
		\begin{split}
		J_n(A_1 \times \cdots \times A_n) :&= p_n \sum_{\sigma \in P_n} \Pi_n(A_{\sigma(1)}\times \cdots \times A_{\sigma(n)}) \\
		&= n! p_n \Pi^{sym}_n (A_1 \times \cdots \times A_n).
		\end{split}
		\end{equation*}
	\end{definition}
	From the normalisation condition \eqref{JM-normcond} we can see that $J_n$ is not a probability measure. In fact, observing that $\Pi_n(\X^{(n)}) = 1$, we can write
	\begin{equation}\label{JM-normcondJ}
	\sum_{n = 0}^{\infty} \frac{J_n(\X^{(n)})}{n!} = 1,
	\end{equation}
	where we interpret $J_0(\X^{(0)}) = p_0$. For any $n \geqslant 1$ we have 
	\begin{equation}\label{JM - nprob}
	J_n(\X^{(n)}) = p_n n!
	\end{equation}
	It is clear that any family of symmetric measures fulfilling the normalisation condition \eqref{JM-normcondJ}, can be used to construct the probability distribution $\{p_n\}_{n = 0}^\infty$, using \eqref{JM - nprob}, and so also the sets of symmetric probability distributions $\{\Pi^{sym}_n\}_{n=0}^\infty$. Now we want to show explicitly how to construct the fidis of the point process from the Janossy measures. In order to do that, we recall that the multinomial coefficient
	\begin{equation*}
	\left( \begin{split}
	& n \\
	n_1,\cdot &\cdot \cdot ,n_r
	\end{split}        
	\right) = \frac{n!}{n_1!\cdots n_r!}
	\end{equation*}
	counts the number of ways we may arrange $n = n_1 + \cdots + n_r$ objects in $r$ different boxes putting $n_1$ objects in the $1$-th box, \dots, $n_r$ objects in the $r$-th box. This implies that
	\begin{widetext}
		\begin{equation}\label{JM-firstformula}
		\begin{split}
		\mu_{\xi} (A_1, n_1; \cdots; A_r, n_r) = 
		p_n\left( \begin{split}
		& n \\
		n_1,\cdot &\cdot \cdot ,n_r
		\end{split}    \right) \Pi_n^{sym}(A_1^{(n_1)} \times \cdots \times A_r^{(n_r)}) = \frac{J_n(A_1^{(n_1)} \times \cdots \times A_r^{(n_r)})}{n_1! \cdots n_r!},
		\end{split}
		\end{equation}
	\end{widetext}
	where $A_1, \cdots A_n$ form a finite partition on $\X$. In general, the sets on which the fidis can be evaluated do not form a partition of $\X$, however we can still find them by the Janossy measure. Suppose that $A_1,\cdots, A_r$ are disjoint sets but they do not from a partition. Hence there exists a set $C= (A_1 \cup \cdots \cup A_r)^c$ such that $\X = A_1 \cup\cdots \cup A_r \cup C$  and clearly $A_i \cap C = \{\varnothing\}$. Thus $A_1,\cdots,A_r,C$ is a partition of $\X$ and the previous formula applies. Let $n = n_1 + \cdots + n_r$ and $s \in \Nat$, we can write that
	\begin{equation*}
	\begin{split}
	\mu_{\xi} (A_1, n_1; & \cdots; A_r, n_r;C,s) = \\
	&\frac{J_{n+s}(A_1^{(n_1)} \times \cdots \times A_r^{(n_r)} \times C^{(s)})}{n_1! \cdots n_r!s!}.
	\end{split}
	\end{equation*}
	Using the law of total probability, we have
	\begin{equation*}
	\mu_{\xi}(A_1,n_1; \cdots; A_r,n_r) = \sum_{s = 0}^{\infty}\mu_{\xi} (A_1, n_1; \cdots; A_r, n_r;C,s)
	\end{equation*}
	and so
	\begin{equation}\label{JD-almostfidi}
	\begin{split}
	\mu_{\xi}(A_1,n_1; &\cdots; A_r,n_r) = \\
	& \frac{1}{n_1!\cdots n_r!}\sum_{s = 0}^\infty \frac{J_{n+s}(A_1^{(n_1)} \times \cdots \times A_r^{(n_r)} \times C^{(s)})}{s!}.
	\end{split}
	\end{equation}
	At this point is should be clear that Janossy measures are an important tool to describe a point process. However nothing was said about its meaning. We already observed that they are not probability measures, thus a statistical interpretation is not available in general. By the way, in some particular case such interpretation is available.
	\begin{definition}
		Let $\xi: (\Omega,\E,P) \rightarrow (\mathcal{N}_{\X}, \mathscr{B}(\mathcal{N}_{\X}))$ be a point process on $\X$ and let $J_n(A_1 \times \cdots \times A_n)$ be a $n$-th Janossy measure of the process. Let $\mu$ be a Borel measure on $\X$. The function $j_n(x_1,\cdots,x_n)$ such that
		\begin{equation*}
		\begin{split}
		J_n( A_1 \times & \cdots \times A_n) = \\
		&\int_{A_1}\cdots \int_{A_n}j_n(x_1,\cdots,x_n)\mu(dx_1) \cdots \mu(dx_n)
		\end{split}
		\end{equation*}
		is called ($n$-th) \emph{Janossy density} of the point process $\xi$ with respect to $\mu$.
	\end{definition}
	The Janossy densities have a particularly simple interpretation, in fact $j_n(x_1,\cdots,x_n)\mu(dx_1)\cdots \mu(dx_n)$ represent the \emph{probability that there are exactly $n$ points in the process $\xi$, one in each of the $n$ distinct infinitesimal regions $(x_i,x_i + dx_i)$}. The notion of Janossy measure (or density, if it exist) can be extended to the case of locally finite point processes by restricting the point process $\xi$ on some compact region $\Lambda \subset \X$. Everything remains the same except that the normalisation condition \eqref{JM-normcondJ} and \eqref{JM - nprob} are not computed with $\X^{(n)}$ but by using $\Lambda^{(n)}$. Similarly in the equation \eqref{JD-almostfidi}, $A_1, \cdots, A_r$ are a partition of $\Lambda$ not of $\X$. For this reason, it is typical to make this $\Lambda$-dependence explicit by adding a label in the symbol of the Janossy measure (or density), i.e. $J_n(A_1 \times \cdots \times A_n | \Lambda)$ and $j_n(x_1.\cdots,x_n | \Lambda)$ which are called \emph{local Janossy measure} and \emph{density}, respectively. The existence of a the Janossy density when $\X =\Rea^d$ and the measure $\mu$ is the ordinary Lebesgue measure, can be used to define an important class of point processes.
	\begin{definition}
		Let $\xi$ be a point process on $\X \subseteq \Rea^d$. If, for all $n \geqslant 1$ and some bounded $A \in \borell{d}$, the local Janossy measures $J_n(dx_1 \times \cdots \times dx_n | A)$ exist and are absolutely continuous with respect to the Lebesgue measure, the point process is said \emph{regular on $A$}. If this happens for any bounded $A \in \borell{d}$, then $\xi$ is said \emph{regular point process}.
	\end{definition}
	Regularity is important because it implies \emph{simpleness}: more precisely, if a point process $\xi$ on $\Rea^d$ has Janossy measure admitting Janossy densities with respect to the Lebesgue measure, then $\xi$ is simple (see Prop. 5.4 V in \cite{daley2007introduction}).
	
	Let us now describe the second quantity which is typically used to describe a point process: the moment measure. By definition, a point process is described by using \emph{random measures}, i.e. measure-valued random variables. Moment measures are just the moments of these random variables, which are measure-valued. This description of a point process $\xi$ on $\X$, is clearly related to the description via the Janossy density. Let us start with the simplest case: the \emph{intensity measure} or \emph{1st-moment measure}. The intensity measure of a locally finite point process $\xi$ in $\X$ is defined as the the measure
	\begin{equation}\label{1th-mommes}
	M_1(A) := \Ex \bigg[ \sum_{x \in \xi} \chi_{A}(x) \bigg],
	\end{equation}
	where $\chi_A$ is the indicator function of the set $A \in \borelx$. The intensity measure is by definition the expectation value of the counting measure on $A$, i.e. $M_1(A) = \Ex[N_\xi(A)]$, and so it can be interpreted as the expected number of points of $\xi$ in $A$. Similarly one can define the \emph{$2$nd-moment measure} as
	\begin{equation}\label{2th-mommes}
	M_2(A_1 \times A_2) := \Ex \bigg[ \sum_{x,y \in \xi} \chi_{A_1 \times A_2}(x,y) \bigg].
	\end{equation}
	Also in this case, we can conclude that $M_2(A_1 \times A_2) = \Ex[N_\xi(A_1)N_\xi(A_2)]$. Recognising that $N_{\xi}(A_1)N_{\xi}(A_2)$ is the number of elements of the set $\{ (x,y) \in A_1 \times A_2, x,y \in \xi \}$, we can interpret the $2$-th moment measure as the intensity measure of a point process on $\X \times \X$. In this definition  of $M_2$, two contributions can be distinguished:
	\begin{widetext}
		\begin{equation*}
		\begin{split}
		M_2(A_1 \times A_2) &= \Ex \bigg[ \sum_{x,y \in \xi} \chi_{A_1 \times A_2}(x,y) \bigg] = \Ex \bigg[ \sum_{\substack{x,y \in \xi \\  x \neq y}} \chi_{A_1 \times A_2}(x,y) \bigg] + \Ex \bigg[ \sum_{x \in \xi} \chi_{A_1 \times A_2}(x,x) \bigg] \\
		&= \Ex \bigg[ \sum_{\substack{x,y \in \xi \\  x \neq y}} \chi_{A_1 \times A_2}(x,y) \bigg] + M_1(A_1 \cap A_2),
		\end{split}
		\end{equation*}
	\end{widetext}
	where we used the fact that $\chi_{A_1 \times A_2}(x,x)$ is non zero only for $x \in A_1 \cap A_2$, i.e. $\chi_{A_1 \times A_2}(x,x) = \chi_{A_1 \cap A_2}(x)$. The quantity
	\begin{equation}\label{2th-facmommes}
	M_{[2]}(A_1 \times A_2) := \Ex \bigg[ \sum_{\substack{x,y \in \xi \\ x \neq y}} \chi_{A_1 \times A_2}(x,y) \bigg]
	\end{equation}
	is called \emph{2nd-factorial moment}. Like for the 2nd moment measure, the 2nd factorial moment measure can be seen as the intensity measure of a point process on $\X \times \X$ consisting of all the $2$-tuples of distinct points of the original process $\xi$. Note that
	\begin{equation*}
	\begin{split}
	M_{[2]}(A \times A) &= M_2(A \times A) - M_1(A) \\
	&= \Ex[N_\xi(A)^2] - \Ex[N_\xi(A)] \\
	&= \Ex[N_{\xi}(A)(N_\xi(A) - 1)].
	\end{split}
	\end{equation*}
	The moment measures defined by \eqref{1th-mommes}, \eqref{2th-mommes} and \eqref{2th-facmommes} can be generalised as follows.
	\begin{definition}\label{MM-def}
		Let $\xi$ be a point process on $\X$, $n \in \Nat$ and $A_1,\cdots,A_n \in \borelx$. The measure
		\begin{equation*}
		M_n(A_1 \times \cdots \times A_n) := \Ex \bigg[ \sum_{x_1, \cdots, x_n\in \xi} \chi_{A_1 \times \cdots \times A_n} (x_1,\cdots, x_n)\bigg]
		\end{equation*}
		is said \emph{$n$-th moment measure} of the process $\xi$, while the measure
		\begin{equation*}
		M_{[n]}(A_1 \times \cdots \times A_n) := \Ex \bigg[ \sum_{\substack{
				x_1, \cdots, x_n\in \xi \\ 
				x_1 \neq \cdots \neq x_n
			}} \chi_{A_1 \times \cdots \times A_n} (x_1,\cdots, x_n)\bigg]
			\end{equation*}
			is said \emph{$n$-th factorial moment measure} of the process $\xi$.
		\end{definition}
		In general, these measures may not exist for any $n$ (they can be infinte sometimes). The $n$-th moment measure can be written as the expectation of a product of counting measures. More precisely, let $A_1,\cdots,A_r \in \borelx$ with $r \leqslant n$, then  
		\begin{equation*}
		M_n(A_1^{(n_1)}\times \cdots \times A_r^{(n_r)}) = \Ex[(N_\xi(A_1))^{n_1}\cdots (N_\xi(A_r))^{n_r}]
		\end{equation*}
		where $n_i \in \Nat$ for all $i = 1, \cdots, r$ and $n_1 + \cdots + n_r = n$. A similar formula for the $n$-th factorial moment is not available in general, but a similar result holds for disjoint sets. Indeed, introducing the \emph{$r$-th factorial power of $x$}
		\begin{equation*}
		x^{[r]} := \begin{cases}
		x(x-1)\cdots(x-r+1) &\mbox{ for }  r \leqslant x \\
		0 &\mbox{ otherwhise}.
		\end{cases}
		\end{equation*}
		where $x \in \Rea$ and $r \in \Nat$, if $A_1,\cdots,A_r \in \borelx$ with $r \leqslant n$ are disjoint sets, one can write that
		\begin{equation}\label{kth-facmommes-dis}
		M_{[n]}(A_1^{(n_1)}\times \cdots \times A_r^{(n_r)}) = \Ex[(N_\xi(A_1))^{[n_1]}\cdots (N_\xi(A_r))^{[n_r]}]
		\end{equation}
		where again $n_i \in \Nat$ for all $i = 1, \cdots, r$ and $n_1 + \cdots + n_r = n$.\newline
		
		Moment measures and the Janossy measures are related. In particular, we can pass from the Janossy measures to the moment measures using
		\begin{equation}\label{MM-JM}
		\begin{split}
		M_{[n]}( & A_1^{(n_1)}\times \cdots \times A_r^{(n_r)}) = \\
		&\sum_{s = 0}^\infty \frac{J_{n+s}(A_1^{(n_1)} \times \cdots \times A_r^{(n_r)} \times \X^{(s)})}{s!}.
		\end{split}
		\end{equation}
		This relation can be inverted provided that all the moments exist (i.e. $M_{[n]}(\X^{(n)}) < \infty$ for any $n \in \Nat$), getting
		\begin{equation}\label{JM-MM}
		\begin{split}
		J_{n}( & A_1^{(n_1)}\times \cdots \times A_r^{(n_r)}) = \\
		&\sum_{s = 0}^\infty (-1)^s \frac{M_{[n+s]}(A_1^{(n_1)} \times \cdots \times A_r^{(n_r)} \times \X^{(s)})}{s!}.
		\end{split}
		\end{equation}
		Also in this case it is not necessary that the sets $A_1,\cdots,A_r$ form a partition of $\X$. If the point process is not finite, these last two equations may still be used by replacing $\X$ with some subset $\Lambda \subset \X$ and by using the local Janossy measures on $\Lambda$. Also in this case these measures, $M_n$ or $M_{[n]}$, may admit densities with respect to some measure  $\mu$ on $\X$.
		\begin{definition}
			Consider a point process $\xi$ and a measure $\mu$, both defined on $\X$. If, given $A_1, \cdots, A_r \in \borelx$ disjoint subsets of $\X$ and $n_i \in \Nat$ for all $i = 1, \cdots, r$ such that $n_1 +\cdots + n_r = n$, we can write
			\begin{equation*}
			\begin{split}
			M_{[n]}( & A_1^{(n_1)} \times \cdots \times A_r^{(n_r)}) \\
			&= \int_{A_1^{(n_1)} \times \cdots \times A_r^{(n_r)}} \rho_n(x_1,\dots,x_n) \mu(dx_1) \cdots \mu(dx_n)
			\end{split}
			\end{equation*}
			the function $\rho(x_1,\dots,x_n)$ is said \emph{$n$-th correlation function} of the process $\xi$.
		\end{definition}
		It is not difficult to see that the relations \eqref{MM-JM} and \eqref{JM-MM} can be used to relate these densities with the Janossy densities and viceversa.  Also in this case we have an interpretation of the quantity $\rho(x_1, \cdots, x_n) \mu(dx_1) \cdots \mu(dx_n)$: it represents \emph{the probability to find  at least $n$ points of the process $\xi$, one in each of the $n$ distinct infinitesimal regions $(x_i, x_i + dx_i)$ }. In contrast with the Janossy densities, in this case the number of points in the intervals is not fixed (for $j_n(x_1, \cdots, x_n)$ the number is \emph{exactly} $n$). Note that \eqref{JM-MM} and \eqref{MM-JM} can be used to pass from Janossy to moment densities and viceversa. \newline
		
		\subsection{Marked point processes}\label{sec1.2}
		
		A very useful extension of the notion of point process is the one of \emph{marked point process}, whose main features will be briefly presented here. The main references for this section remains \cite{baddeley2007spatial} and \cite{daley2007introduction}.\newline
		
		There can be situations where the point process is not the principal object that one wants to analyse, like for example, when the point process is just a component of a more complex model. In such situations, it is sometimes useful to associate to each point $x_i$ of the point process, an additional variable $m_i$ belonging to some set $\M$. Such variables are called \emph{marks} and can be anything:  for example they can be a label (e.g. the time at which an event happened), a random variable or a set. The set $\M$ containing all marks is said \emph{mark space}. The resulting point process, whose generic points are represented by the couple $(x_i,m_i)$, is said \emph{marked point process}. 
		\begin{definition}
			A \emph{marked point process} on $\X$ is a point process on $\X \times \M$ having points $\tilde{\xi}=\{(x_1,m_1), (x_2,m_2), \cdots\}$, such that $N_g(A):=N_{\tilde{\xi}}(A \times \M) < \infty$ for any $A \in \borelx$. The point process $\xi_g$ defined with the measures $\{N_g(A)\}_{A \in \borelx}$ is said \emph{ground process}.
		\end{definition}
		Not all point processes on the product space $\X \times \M$ are marked point processes, but only those for which the ground process is still a point process. A rather simple case when this is always possible is when the mark space is a finite set, i.e. $\M := \{1, \cdots, k\}$ for some $k \in \Nat$. The marked point process in this case is said \emph{multivariate}, and the finiteness condition of the marked point process is always satisfied. In fact, we have
		\begin{equation}\label{MPP - CountM}
		N_g(A) = N_{\tilde{\xi}}(A \times \{1,\cdots,k\}) = \sum_{i = 1}^m N_{\tilde{\xi}}(A \times \{i\}),
		\end{equation} 
		which follows from the additivity of counting measures between disjoint sets (note that $(A \times \{i\}) \cap (A \times \{j\}) = \{\varnothing\}$ for $i \neq j$). Since $N_{\tilde{\xi}}(A \times \{i\}) < \infty$ for all $A \in \borelx$ then also $N_g(A) < \infty$ for all $A \in \borelx$, showing that if the space of marks is a finite set, then any point process on $\X \times \M$ is marked. The counting measures $N_i(A):=N_{\tilde{\xi}}(A \times \{i\})$ define point processes on $\X \times \{i\}$, for any $i \in \M$, which are sometimes called \emph{component processes} of $\tilde{\xi}$.
		
		\subsection{Determinantal point processes}\label{sec1.3}
		
		Here we want to describe an interesting class of point processes: the determinantal point processes. Our interest in this particular class of processes is mainly due to the fact that the whole statistical properties are determined by the kernel of a (locally) trace-class operator over a separable infinite-dimensional Hilbert space. The main references are \cite{macchi1975coincidence,hough2009zeros,hough2006determinantal,decreusefond2016determinantal,soshnikov2000determinantal,shirai2003random}, while  a brief explanation of the main mathematical notions used (like integral kernels, locally trace-class operators and Fredholm's determinants) is given in appendix A.\newline
		
		From now on, we assume $\X = \Rea^d$ for simplicity. Let $\hat{K}$ be an operator acting on $L_2(\Rea^d)$ such that
		\begin{enumerate}
			\item[As1)] $\hat{K} \in \Ltco \cap \mathcal{B}_2(L_2(\Rea^d)$, namely $\hat{K}$ admits kernel $K(x,y)$ and 
			\begin{equation*}
			Tr(\hat{P}_{\Lambda}\hat{K}\hat{P}_{\Lambda}) < \infty 
			\end{equation*}
			where $\Lambda \subset \Rea^d$ is compact and $\hat{P}_{\Lambda}: L_2(\Rea^d) \rightarrow L_2(\Lambda)$ is a projector;
			\item[As2)] $\hat{\mathbb{O}} \leqslant \hat{K} < \Id$, namely the spectrum of $\hat{K}$ is in $[0,1)$;
			\item[As3)] $\hat{K}$ is a self-adjoint operator, which implies that the associated kernel $K(x,y) : \Rea^d \times \Rea^d \rightarrow \Comp$ is hermitian, namely such that
			\begin{equation*}
			K(x,y) = [K(y,x)]^*
			\end{equation*}
			for any $x,y \in \Rea^d$.
		\end{enumerate}
		Given such operator $\hat{K}$, consider its \emph{local version} on $\Lambda \subset \Rea^d$, i.e. $\hat{K}_{\Lambda} := \hat{P}_{\Lambda}\hat{K}\hat{P}_{\Lambda}$. By the Mercer theorem (see Th. \ref{apA:mercer-theorem} in appendix A) the kernel associated to $\hat{K}_{\Lambda}$ can be written as
		\begin{equation}\label{DPP: locally-exp}
		K_{\Lambda}(x,y) = \sum_{{\mu_\Lambda}} {\mu_\Lambda} \varphi_{\mu_\Lambda}(x) \varphi_{\mu_\Lambda}^*(y) 
		\end{equation}
		where $\mu_{\Lambda} \in \Rea$ are eigenvalues and $\varphi_{\mu_\Lambda} \in L_2(\Lambda)$ are the corresponding eigenvectors of $\hat{K}_{\Lambda}$. Note that, in any case, $K_{\Lambda}(x,y) = \chi_{\Lambda}(x)K(x,y)\chi_{\Lambda}(y)$. Now we are ready to define the point process we are interested in.
		\begin{definition}
			Let $\xi$ be a locally-finite simple point process on $\Rea^d$ having, for any $n\in \Nat$, $n$-th factorial moment density given by
			\begin{equation*}
			\rho_n(x_1,\cdots,x_n) = \det( [K(x_i,x_j)]_{i,j = 1 \cdots n}),
			\end{equation*}
			where $x_1,\cdots,x_n \in \Rea^d$ and $K(x,y)$ is the kernel of an operator $\hat{K}$ on $L_2(\Rea^d)$ fulfilling $As1) - As3)$. The point process $\xi$ is said  \emph{determinantal point process (DPP)} on $\Rea^d$.
		\end{definition}
		In the above definition, the writing $[K(x_i,x_j)]_{i,j = 1 \cdots n}$ is a short hand notation for the $n \times n$ matrix whose $(i,j)$-th element is $K(x_i,x_j)$. By construction, in a DPP all the moment densities are well defined. 
		This implies that all the (local) Janossy densities exist, are well defined and can be found using \eqref{MM-JM}. Following \cite{decreusefond2016determinantal}, we derive the Janossy density in a different way. Given $\hat{K}$ in $L_2(\Rea^d)$, operator whose kernel define a DPP on $\Rea^d$, let us define
		\begin{equation*}
		\hat{J}_{\Lambda} := (\Id - \hat{K}_{\Lambda})^{-1} \hat{K}_{\Lambda}.
		\end{equation*}
		This operator is usually called \emph{local interaction operator}. It can be proved that $\hat{J}_{\Lambda} \in \mathcal{B}_1(L_2(\Rea^d))$, hence it admits kernel, denoted by the symbol $\mathcal{J}_\Lambda(x,y)$, and in particular from \eqref{DPP: locally-exp}, we can write that
		\begin{equation*}
		\mathcal{J}_\Lambda(x,y) = \sum_{\mu_{\Lambda}} \frac{\mu_{\Lambda}}{1 - \mu_{\Lambda}} \varphi_{\mu_\Lambda}(x) \varphi_{\mu_\Lambda}^*(y).
		\end{equation*}
		Given a configuation of the DPP, say $\{x_1,\cdots,x_n\}$, define the function
		\begin{equation}\label{DPP-localinteractiondet}
		\eta_{\Lambda}(x_1,\cdots,x_n) := \det( [\mathcal{J}_\Lambda(x_i,x_j)]_{i,j =1,\cdots,n} ).
		\end{equation}
		Then the Janossy densities can be computed using the following result \cite{decreusefond2016determinantal,shirai2003random}.
		\begin{proposition}
			Let $\hat{K}$ be an operator on $L_2(\Rea^d)$ fulfilling $As1) - As3)$ and $K(x,y)$ its associated kernel. Let $\xi$ be a DPP on $\Rea^d$ with kernel $K(x,y)$. Then for all compact subset $\Lambda \subset \Rea^d$ and $n \in \Nat/\{0\}$, the local Janossy density of the process $\xi$ is
			\begin{equation*}
			j_n(x_1,\cdots,x_n|\Lambda) = \det (\Id - \hat{K}_{\Lambda} ) \eta_{\Lambda}(x_1,\cdots,x_n)
			\end{equation*}
			while $j_0(\Lambda) = \det (\Id - \hat{K}_{\Lambda} )$.
		\end{proposition}
		In the above proposition, $\det (\Id - \hat{K}_{\Lambda} )$ must be interpreted as Fredholm determinant (see appendix A).
		Recalling the discussion done in the section \ref{sec1.1}, the kernel $K(x,y)$ of a DPP $\xi$ contains all the information needed to completely characterise the process. Indeed, it allows to compute all the fidis of $\xi$, but we may also deduce other properties. Moreover, from the $1$st moment measure, for any $A \subset \Rea^d$ we have that
		\begin{equation*}
		\begin{split}
		\Ex[N_{\xi}(A)] &= \int_A K(x,x)dx \\
		&= \int_{\Rea^d} \chi_A(x) K(x,x)dx \\
		&= Tr(\hat{P}_A \hat{K}) = Tr(\hat{K}_A)
		\end{split}
		\end{equation*}
		where $\hat{P}_A: L_2(\Rea^d) \rightarrow L_2(A)$ is an orthogonal projector. Setting $A = \Rea^d$, we can see that the expected number of points is nothing but that trace of $\hat{K}$. This suggests that directly from $\hat{K}$, we can obtain information on the finiteness of the DPP. This theorem formalise exactly this idea \cite{decreusefond2016determinantal,soshnikov2000determinantal}.
		\begin{theorem}\label{DPP-Noptheo}
			Let $\hat{K}$ be an operator on $L_2(\Rea^d)$ fulfilling $As1) - As3)$ and let $\xi$ the DPP generated by the associated kernel. We have
			\begin{enumerate}
				\item[i)] If $Tr(\hat{K}) < \infty$, i.e. $\hat{K} \in \mathcal{B}_1(L_2(\Rea^d))$ then $P[N_\xi(\Rea^d) < \infty] = 1$ which means that the point process is finite with probability 1, while when $Tr(\hat{K}) = \infty$ the point process is only locally-finite, namely $P[N_\xi(\Rea^d) < \infty] = 0$;
				\item[ii)] $P[N_\xi(\Rea^d) \leqslant m] = 1$ for some $m \in \Nat/\{0\}$ if and only if $\hat{K}$ has finite rank and $\mbox{Rank}(\hat{K}) \leqslant m$;
				\item[iii)] The number of points of the process is exactly $m \in \Nat/\{0\}$ if and only if $\hat{K}$ is an orthogonal projector having $\mbox{Rank}(\hat{K}) = m$.
			\end{enumerate}
		\end{theorem}
		A DPP whose kernel is an orthogonal projector is said \emph{orthogonal}. Finally, we conclude by explaining what happens to a DPP when we remove a point. For a point process, this operation is called \emph{thinning}. DPPs are closed under thinning in the sense that the new point process obtained after this operation is still a DPP. More formally, given a DPP $\xi$ on $\Rea^d$ we may form a new point process as follows. Given a point $z \in \Rea^d$, we condition the DPP on the event $z \in \xi$ and then remove this point. With this procedure a new point process is obtained, say $\xi/\{z\}$, which is the DPP $\xi$ without the point $z$. For the DPP obtained with this procedure the following theorem holds\cite{shirai2003random,Kui}.
		\begin{theorem}\label{DPP-thin}
			Let $\xi$ be a DPP with kernel $K(x,y)$ and $z \in \Rea^n$ such that $0<K(z,z)<\infty$. Then the point process obtained by removing $z$ from $\xi$ as described above is again a DPP with kernel $\tilde{K}(x,y)$ given by
			\begin{equation*}
			\tilde{K}(x,y) = K(x,y) - \frac{K(x,z)K(z,y)}{K(z,z)}.
			\end{equation*}
		\end{theorem}
		This theorem can be extended also to points $z \in \Rea^d$ for which $K(z,z) = 0$ or $K(z,z) = \infty$, if we can approximate $z$ with points $s$ all having $0 < K(s,s) < \infty$. Also in this case the process $\xi / \{z\}$ is still a DPP with kernel given by
		\begin{equation*}
		\tilde{K}(x,y) = K(x,y) - \lim_{s \rightarrow z}\frac{K(x,s)K(s,y)}{K(s,s)}.
		\end{equation*}
		
		\subsection{DPP diffusion}\label{sec1.4}
		
		In the description of the DPP done in section \ref{sec1.3}, we did not mention time at any level. In this sense the DPP (as any point process) can be considered as a \textquotedblleft spatial process", i.e. a collection of random variables parametrised by space (see definition \ref{PP-generaldef}). In ordinary stochastic processes, the parameter is typically assumed to be the time and time-evolution is stochastic. Here we want to describe, at a very qualitative level, how it is possible to implement a stochastic time-evolution for a DPP. In particular, we will describe how to implement a \emph{diffusive} time-evolution: this will be done by using suitable \emph{Dirichlet forms}.
		
		Let us start with an analogy. The diffusion process we are going to describe can be imagined as follow. A DPP can be thought as a classical gas of fermions\footnote{The word \textquotedblleft classical" in this context should be interpreted as follows: a gas of \emph{classical} fermion is a gas of particles having $n$-point correlation functions equal to the one of a quantum gas of fermions. However, no quantum description of these particles is needed for the definition of such a gas  \cite{macchi1975coincidence,soshnikov2000determinantal}. } and the diffusion process we want to describe is the diffusion of such a gas. Note that, because of the fermionic character, the gas cannot be thought as non-interacting: even by neglecting all the possible interactions between particles two fermions cannot be found in the same place, i.e. there is an exchange-correlation interaction to be taken into account. Such interaction prevents the formation of clusters in the process: in this sense, it is a repulsive interaction between particles. A possible approach to describe this diffusion process can be the following. Given a DPP configuration $\xi$ one marks each point with a label $t \in \Rea^+$. The $t$-components of marked point process thus obtained, $\xi_t= \xi \times \{t\}$, is made of couples $(x_i,t)$, with $x_i \in \xi$ and $t \in \Rea^+$, representing the spatial and temporal location of the point. Then we can imagine that each point in $\xi_t$ evolves in time according to some stochastic differential equation (SDE), one for each point. Since we have an infinite number of points (in general), we have an infinite number of SDEs. In addition, to model the repulsiveness due to the fermionic character of the gas, the SDEs describing the trajectories of each particle should have a drift term which depends on the locations of all the other particles. Thus, if we want to find the configuration of this gas (the position of each particle) at a given time $t \in \Rea^+$ (knowing its configuration at some time $t_0 < t$), we should solve an infinite system of coupled SDEs. It is not easy to do so, however this problem can be tackled from a different perspective using the so called Dirichlet forms, as explained in \cite{osada1996dirichlet}. In appendix $B$, the general notion of Dirichlet forms and its connection with Markov processes is briefly reviewed.
		
		The Dirichlet forms approach for the diffusion of a DPP is used in \cite{yoo2005dirichlet} for a particular class of DPP, those having translational invariant kernels. The application of the same method to more general cases, can be found in \cite{decreusefond2016determinantal} and \cite{decreusefond2012stochastic}: here we report the main steps explained in these works, for the construction of such a diffusion. By definition of point process, a DPP is a random variable taking values on the space $(\mathcal{N}_{\Rea^d}, \mathscr{B}(\mathcal{N}_{\Rea^d}))$. Let $\mu_{dpp}$ be the probability distribution for such a process. If $\xi$ is a configuration of the DPP, take its restriction to $\Lambda$, compact subset of $\Rea^d$, i.e. $\xi_\Lambda := \xi \cap \Lambda$. Then define $\mathcal{N}_{\Lambda}^f$ as the subset of $\mathcal{N}_{\Rea^d}$
		\begin{equation*}
		\mathcal{N}_{\Lambda}^f := \{ N' \in \mathcal{N}_{\Rea^d}| N'(\Lambda) < \infty\},
		\end{equation*}
		i.e. the set of all the \emph{finite} simple counting measures on $\Lambda$, which is the set on which $\xi_{\Lambda}$ takes values. To the restrincted DPP $\xi_\Lambda$, one can associate an Hilbert space $L_2(\mathcal{N}_{\Lambda}^f,\mu_{dpp})$. Consider the set of real valued functions $F: \mathcal{N}_{\Lambda}^f \rightarrow \Rea$, which can be written as
		\begin{equation*} 
		F(\xi_\Lambda) = f_0 \chi_{\{N_{\xi}(\Lambda) = 0\}} + \sum_{i =1}^{n} \chi_{\{N_{\xi}(D) = i\}} f_i(x_1,\cdots, x_i)
		\end{equation*}
		where $x_1,\cdots,x_n \in \xi_{\Lambda}$, $n\geq 1$ is an integer, $f_0$ a constant and $f_i$ are smooth symmetric functions. Call this set $\mathcal{S}_{\Lambda}$. It can be proved that $\mathcal{S}_{\Lambda}$ is dense in $L_2(\mathcal{N}_{\Lambda}^f,\mu_{dpp})$, hence $F(\xi_\Lambda)$ can seen as the typical element of $L_2(\mathcal{N}_{\Lambda}^f,\mu_{dpp})$. On $\mathcal{S}_{\Lambda}$, one can define a gradient
		\begin{equation*}
		\nabla_{x}^{\mathcal{N}_{\Lambda}^f} F(\xi) := \sum_{k = 1}^n \chi_{\{N_{\xi}(\Lambda) = k\}} \sum_{y \in \xi} \chi_{\{y = x\}} \nabla_x f_k(x_1,\cdots,x_k),
		\end{equation*}
		where $\nabla_x$ is the ordinary gradient in $\Rea^d$. This last definition allows to introduce the following symmetric real bilinear form on $\mathcal{S}(\Lambda)$
		\begin{equation*}
		\varepsilon_{\Lambda}(F,G) = \Ex \bigg[ \sum_{y \in \xi_{\Lambda}} \nabla_{y}^{\mathcal{N}_{\Lambda}^f}F(\xi_\Lambda) \cdot \nabla_{y}^{\mathcal{N}_{\Lambda}^f} G(\xi_{\Lambda}) \bigg],
		\end{equation*}
		where $\cdot$ denotes the ordinary scalar product in $\Rea^d$. This form can be extended to the whole $L_2(\mathcal{N}_{\Lambda}^f,\mu_{dpp})$ and, under suitable conditions, can be rewritten as (see lemma 4.2, \cite{decreusefond2012stochastic})
		\begin{equation*}
		\Ex \bigg[ \sum_{y \in \xi_{\Lambda}} \nabla_{y}^{\mathcal{N}_{\Lambda}^f}F(\xi_\Lambda) \cdot \nabla_{y}^{\mathcal{N}_{\Lambda}^f} G(\xi_{\Lambda}) \bigg] =
		\Ex [F(\xi_{\Lambda}) \hat{H}_{\Lambda} G(\xi_{\Lambda})],
		\end{equation*}
		where $\hat{H}_{\Lambda}$ is a symmetric non-negative definite operator acting on $L_2(\mathcal{N}_{\Lambda}^f,\mu_{dpp})$. Considering the closure of such operator, one obtains the from
		\begin{equation}\label{DPPdif - diri}
		\overline{\varepsilon}_{\Lambda}(F,G) = \Ex[F(\xi_{\Lambda}) \overline{\hat{H}_{\Lambda}} G(\xi_{\Lambda})].
		\end{equation}
		Then $(\overline{\varepsilon}_{\Lambda}, \mbox{Dom}(\overline{\hat{H}_{\Lambda}^{1/2}}))$ is a Dirichlet form on $L_2(\mathcal{N}_{\Lambda}^f,\mu_{dpp})$ (see Th. 4.1, \cite{decreusefond2012stochastic}). This Dirichlet form, which is defined on the space $\mathcal{N}_{\Lambda}^f$ on which the DPP $\xi_{\Lambda}$ takes values, can be used to construct a stochastic time-evolution for $\xi_{\Lambda}$ which is a diffusion. In particular the following theorem holds (Th. 5.1, \cite{decreusefond2012stochastic}).
		\begin{theorem}\label{DPPdif - fundtheo}
			Let $\xi$ be a DPP on $\Rea^d$ with kernel $K(x,y)$ and probability distribution $\mu_{dpp}$. Let $\hat{K}$ the locally trace-class integrable operator associated to this kernel from which one can construct the function $\eta_{\Lambda}(x_1,\cdots,x_n)$ as in \eqref{DPP-localinteractiondet}, where $\Lambda$ is a compact subset of $\Rea^d$. Assuming that
			\begin{enumerate}
				\item[a)] $(x_1,\cdots,x_n) \mapsto \eta_{\Lambda}(x_1,\cdots,x_n)$ is continuously differentiable on $\Lambda^n$;
				\item[b)]  for any $n \in \Nat$, $1 \leqslant i,j \leqslant n$ and $1 \leqslant h,k \leqslant d$,
				\begin{equation*}
				\begin{split}
				\int_{\Lambda^n} &\bigg| \frac{\partial_{x_i^{(h)}}\eta_{\Lambda}(x_1,\cdots,x_n) \partial_{x_j^{(k)}} \eta_{\Lambda}(x_1,\cdots,x_n)}{\eta_{\Lambda}(x_1,\cdots,x_n)} \bigg| \cdot \\
				&\cdot \chi_{\{\eta_{\Lambda}(x_1,\cdots,x_n) > 0\}}(x_1,\cdots,x_n) dx_1\cdots dx_n < \infty,
				\end{split}
				\end{equation*}
				where $\partial_{x_i^{(h)}}$ is the derivative with respect to the $h$-th component of $x_i \in \xi$;
			\end{enumerate}
			Then there exists a Markov process $\{\xi_t\}_{t \in \Rea^+}$ on $\mathcal{N}_{\Lambda}^f$ such that
			\begin{enumerate}
				\item[i)] $\{\xi_t\}_{t \in \Rea^+}$ is a diffusion (i.e. the trajectories are continuous with probability one);
				\item[ii)] The transition probabilities of $\{\xi_t\}_{t \in \Rea^+}$ can be calculated using the semigroup $\hat{T}_t$ associated to the Dirichlet from $(\overline{\varepsilon}_{\lambda}, \mbox{Dom}(\overline{\hat{H}_{\Lambda}^{1/2}}))$ defined in \eqref{DPPdif - diri};
				\item[iii)] $\{\xi_t\}_{t \in \Rea^+}$ is unique up to $\mu_{dpp}$-null sets;
				\item[iv)] $\{\xi_t\}_{t \in \Rea^+}$ has $\mu_{dpp}$ invariant measure.
			\end{enumerate}
		\end{theorem}
		Let us explain a bit further the content of this theorem. Given a DPP $\xi$ and taking its restriction on some compact subset $\Lambda$, one can construct a diffusion process from the Dirichlet form \eqref{DPPdif - diri}. If $\hat{H}_{\Lambda}$ is the generator of the Dirichlet form and $A \in \mathscr{B}(\mathcal{N}_{\Rea^d})$ is an event at time $t_0$ with probability $\mu_{dpp}(A,t_0)$, then the probability of the event $A$ at time $t+t_0$, $p(A,t+t_0)$ can be computed as
		\begin{equation*}
		p(A,t + t_0) = e^{ -t \hat{H}_{\Lambda}} \mu_{dpp}(A,t_0).
		\end{equation*}
		The theorem ensures that such diffusion process is unique and also that the stochastic process at time $t + t_0$ is still determinantal, namely
		\begin{equation*}
		p(A,t + t_0)  = \mu_{dpp}(A, t+ t_0).
		\end{equation*}
		Hence, such time-evolution preserves the determinantality of the process. Since the determinantal point process is simple, the trajectories of the single points of $\xi$ do not collide during this evolution. In fact, we have the following theorem (see Th. 5.3, \cite{decreusefond2012stochastic})
		\begin{theorem}
			Under the assumptions of theorem \ref{DPPdif - fundtheo}, for $d \geqslant 2$ the diffusion is non-colliding.
		\end{theorem}
		
		\section{Again model B: 3-D generalisation and point process structure}\label{sec2}
		
		In this section, we consider model B, introduced in \cite{LC2}, under a different point of view. In particular, we describe such a model, using the general theory of point processes described in section \ref{sec1}. This enables us to generalize model B from 1-D to 3-D in a quite straightforward way. The structure of such 3-D generalization will be used later, in section \ref{sec3}, to modify model B and solve the separability problem.
		
		\subsection{Model B: underlying point process structure and generalization to 3D case}\label{sec2.1}
		
		In model B, two are the main ingredients: the space process $\Spb_t$ and the particle process $\mathscr{P}_t = (X_t, V_t(t'))$. Let us start from the space process $\Spb_t$. Since $\Spb_t$ is a random distribution of points over the real line, we should be able to describe it with the theory of point processes presented in the previous section. Suppose we have a collection of $M$ independent Wiener processes taking values on $\Rea$. At a given time $t$, the subset of $\Rea$ formed by all positions of each Wiener process defines a point process. Call $\xi_t$ such point process, without assuming \emph{distinguishability} we can easily conclude that the fidis density of such process are given by
		\begin{equation}\label{FIDISmb}
		\rho_{\xi_t}(x_1,\cdots,x_M) = \frac{1}{M!} \sum_{\sigma \in P_M} \prod_{i = 1}^M \rho_{W_t^{(\sigma(i))}}(x_{i})
		\end{equation}
		where $\sigma$ is a permutation of $\{1,\cdots,M\}$, $P_M$ is the set of all the permutations over this set, and $\rho_{W_t^{(i)}}(x)$ is the probability density of the $i$-th 1-D Wiener process at time $t$. One may note that, if we assume that the Wiener processes are identically distributed, we recover the probability distribution used in model B for $\Spb_t$ for this particular case. However, without this assumption, this formula does not coincide with the one used in model B: the difference lies in the fact that here we are not assuming the distinguishability of the Wiener process. We will come back on this fact soon, but for the moment we observe that, at this level, we are describing \emph{only} the random pattern formed by these points. Given the expression above, one can immediately realise that the $M$-th Janossy density is
		\begin{equation*}
		j_M(x_1,\cdots,x_M;t) = \sum_{\sigma \in P_M} \prod_{i = 1}^M \rho_{W_t^{(\sigma(i))}}(x_{i}).
		\end{equation*}
		This holds for any time $t \in \Rea^+$ hence, the time evolution of the Wiener processes, can be used to obtain the fidis and Janossy densities at any time. Now consider the particle process of model B, and in particular the position random variable $X_t$. In \cite{LC2}, we defined it as
		\begin{equation}\label{MobBPosition}
		X_t(\omega) := \pi_{I_t(\omega)}(\Spb_t(\omega)) - W_t^{i_0}(\omega)
		\end{equation}
		where $\pi_i(\mathbf{x})$ is the projector on the $i$-th coordinate of the $M$-tuple $\mathbf{x}$ and $I_t$ is a random variable taking values on $\{1,\cdots,M\}$, while $W_t^{i_0}(\omega)$ is the position a origin of the reference frame chosen. The position random variable selects one of the points of $\Spb_t(\omega)$, and takes the value of the difference between the point selected and the origin of the reference frame. To express this selection procedure with this writing, the point of the space process must be distinguishable: only in this way one can choose the $i$-th point. Thus, in order to use the definition above starting for the point process $\xi_t$ with fidis given by \eqref{FIDISmb}, we need to distinguish their points. This can be done introducing by an additional label, say $i$, attached to each point and considering the marked point process obtained in this way. More precisely, consider the point process $\xi_t$, call $x_t^{(i)}$ its $i$-th point according to some arbitrary ordering, and consider the set of points 
		\begin{equation*}
		\tilde{\xi}_t = \{(x^{(1)}_t, 1), \cdots, (x_t^{(M)}, M)\}.
		\end{equation*}
		It is a multivariate marked point process on $\Rea$ (by construction, the ground process is exactly $\xi_t$). Using the marked point process $\tilde{\xi}_t $ we can really select a point of $\xi_t$, thus the definition of $X_t$ used makes sense. More precisely we can say that the point process of model B can be described by using the marked point process $\tilde{\xi}_t$: the fidi densities of $\tilde{\xi}_t$ corresponds to the probability densities of $\mathbb{S}^B_t$ given in \cite{LC2}. In this sense $\mathbb{S}^B_t$ is a particular case of the point process described here. Using \eqref{MPP - CountM} we can write
		\begin{equation*}
		N_{\xi_t}(A) = \sum_{i = 1}^M N_{\tilde{\xi}_t} (A \times \{i\})
		\end{equation*}
		where $A \in \borel$. Each component $N_{\tilde{\xi}_t} (A \times \{i\})$ is a point process consisting of a single point. The selection procedure, due to the random variable $I_t$, can be represented considering the $I_t$-th component of $\tilde{\xi}_t$, i.e. $N_{\tilde{\xi}_t}(A \times \{I_t\})$. Defining $N_{X_t}(A) := N_{\tilde{\xi}_t}(A \times \{I_t\})$, the point process $\xi_{X_t}:=\{N_{X_t}(A)\}_{A \in \borell{3}}$ represents the position of the particle at time $t$. We can use the counting measure $N_{X_t}(A)$ to determine where the particle is: $N_{X_t}(A) \neq 0$ only for those $A$ containing the point $X_t = \pi_{I_t}(\tilde{\xi}_t)$, and this can be used to determine the position of the particle. Compared with \eqref{MobBPosition}, we can see that the origin is assumed to be in $0$ at time $t$. This can be done without loss of generality, since to have a non-zero origin one can always perform a suitable translation. From now on we consider the origin always in $0$. Also the distribution of $X_t$ can be derived from such a measure. In order to see it, let us define the intensity measure
		\begin{equation}\label{MB - part dis}
		\mu_{X_t}(A) := M_1(A) = \Ex_{\xi_{X_t}}[N_{X_t}(A)].
		\end{equation}
		Since $\xi_{X_t}$ consists of a single point, clearly $N_{X_t}(\Rea^3) = N_{\tilde{\xi}_t}(\Rea^3 \times \{I_t\}) =1$, which implies that $\mu_{X_t}(\Rea^3) = 1$, i.e. it is a probability measure. This is the probability distribution of the random variable $X_t$, in fact $\mu_{X_t}(A)$ is the expected number of times we find the particle in the set $A \in \borel$, i.e. its probability to be found in $A$. Note that in \eqref{MB - part dis} the expectation is taken with respect to the point process $\xi_{X_t}$, which depends both on the ground process $\xi_t$ (thus on the marked point process $\tilde{\xi}_t$) and the selection random variable $I_t$. Hence we can write
		\begin{equation*}
		\begin{split}
		\mu_{X_t}(A)  &= \int_A \mu_{X_t}(dx) \\
		&= \int_A \int_{\mathcal{N}_{\Rea^d}} \nu_{X_t|\xi_t = \zeta}(dx)\nu_{\xi_t}(d\zeta) \\
		&= \int_A \int_{\tilde{\mathcal{N}}_{\Rea^d}} \mu_{X_t|\tilde{\xi}_t = \sss_t}(dx)\mu_{\tilde{\xi}_t}(d\sss_t)
		\end{split}
		\end{equation*}
		 Thus the position random variable of model B can be described by using suitable components of the multivariate marked point process constructed from a suitable point process. Before going on, let us discuss the relation between the indistinguishability of the points, the need of a labeling, and the arbitrariness of this labeling. We have seen that the point process $\xi_t$ describes the random pattern of the space process but, in order to define the position random variable, we need to introduce an arbitrary labeling among the points (i.e. we need to consider the marked point process $\tilde{\xi}_t$). The arbitrariness of this labeling, needed to define $\tilde{\xi}_t$, may sound strange. However, this situation is a quite common situation. In fact, suppose we have a point $P \in \Rea$. In order to describe it we need a reference frame, namely, we need to specify an origin $O$ and a way to measure the distance between $O$ and $P$. The choice of these two objects is completely arbitrary. The relation between $\xi_t$, $\tilde{\xi}_t$ and the labeling is similar to the one just described. In particular, to define this ordering one may choose an arbitrary point in $\xi_t$, the origin, and use an arbitrary distance function (for example the ones given in appendix B of \cite{LC2}) to label all the other points: the label $i$ of a point $x \in \xi_t$ contains the information needed to describe the point in the correct way with respect to the chosen origin (which is automatically in $0$). Thus changing  the distance function (which is arbitrary) changes the labeling. In this sense $\xi_t$ can be thought as $\Rea$ without specifying a distance function, while $\tilde{\xi}_t$ can be thought as $\Rea$ \emph{equipped with a particular distance function}. In this sense, the need and the arbitrariness of the labeling introduced above should not be considered as a strange thing. 
		
		The generalization to the 3-D case is straightforward: one can replace the densities $\rho_{W_t^{(i)}}(x_{i})$  associated to the 1-D Wiener process at time $t$ with the corresponding densities associated to the 3-D Wiener process. The $M$-th Janossy densities (and so the fidi densities) obtained in this way are associated to a point process in $\Rea^3$, and this holds at any time $t$. In the 3-D case, the particle position is a 3-vector, namely
		\begin{equation} \label{MB-partpro}
		X_t := (X_t^{1},X_t^{2},X_t^{3}).
		\end{equation}
		As for the 1-D case, equation \eqref{MB - part dis} gives us the distribution $\mu_{X_t}(A)$ for some $A \in \borell{3}$. The velocity random variable can be obtained by using \eqref{MB-partpro} in the definition given for model B. Thus, the velocity random variable is the 3-vector
		\begin{equation}\label{MB-3dvel}
		V_t(t') := \frac{ X_{t'} - X_t } {t' - t}.
		\end{equation}
		Then one can proceed as in \cite{LC2} to prove that, after the removal of the space process, when we represent these random variables on a common Hilbert space, the operators associated to each component do not commute, i.e.
		\begin{equation}\label{MB-comm1}
		[\hat{X}_t^i , \hat{V}_t^i (t')] \neq 0,
		\end{equation}
		where $i = 1,2,3$. Since we want to compare it with ordinary quantum mechanics, we also need to check if the components of $X_t$ commute among them. If it is not so, there is no hope to re-obtain ordinary non-relativistic quantum mechanics from any modification of this model. We recall that the non-commutativity is deduced by means of an entropic uncertainty relation, obtained after conditioning on the point process. The strategy used to derive the bound between the sum of $H_{\sss_t}(X_t)$ and $H_{\sss_t}(V_t(t'))$, is based on the fact that the transition probabilities depend on the point process at different times (say $t$ and $t'$). Indeed, we condition the position and velocity random variables to a given realisation $\sss_t$ but $V_t(t')$ depends  also on $\sss_{t'}$ (since it depends on $X_t$ and $X_{t'}$). Thus the conditioning on $\sss_t$ is not sufficient to have a delta-like probability distribution for $V_t(t')$ when this happens for $X_t$: the minimal uncertainty on $V_t(t')$ is bounded by the uncertainty of $\Spb_{t'}$ given $\sss_t$, and this is the origin of the bound in the sum of the two entropies. If we apply this idea to the 3-D generalisation, we cannot derive any bound between the components of $X_t$. In fact given $X_t^i$ and $X_t^j$, with $i \neq j$, their probability distributions can be derived as marginal of $\mu_{X_t}$ and so we can always define a conditional probability like $\mu_{X_t^i | X_t^{j}}$. However $\mu_{X_t^i | X_t^{j}}$ may depend only to a single $\sss_t$, hence after conditioning on such point process configuration, we can always shrink the transition probabilities $\mu_{X_t^i | X_t^{j}}$, to delta-like measures, which implies that $H_{\sss_t}(X_t^{i}) + H_{\sss_t}(X_t^j) \geqslant 0$. Thus
		\begin{equation}\label{MB-comm2}
		[\hat{X}_t^i , \hat{X}_t^j ] = 0,
		\end{equation}
		for all $i,j = 1,2,3$. However, using this commutation relation together with \eqref{MB-partpro} and \eqref{MB-3dvel}, we have to conclude that
		\begin{equation}\label{MB-comm3}
		[\hat{X}_t^i , \hat{V}_t^j (t')] = \frac{1}{t'-t} [\hat{X}^{i}_t,\hat{X}^j_{t'}],
		\end{equation}
		(note the time indices) and 
		\begin{equation}\label{MB-comm4}
		[\hat{V}_t^i(t') , \hat{V}_t^j (t')] = - \frac{1}{(t'-t)^2} \bigg\{ [\hat{X}^{i}_t,\hat{X}^j_{t'}] + [\hat{X}^{i}_{t'},\hat{X}^j_t]\bigg \},
		\end{equation}
		for all $i \neq j$. Notice that in general $[\hat{X}^{i}_t,\hat{X}^j_{t'}] \neq 0$, and $[\hat{X}^{i}_{t'},\hat{X}^j_t] \neq 0$. This concludes the 3-D generalisation of model B presented in \cite{LC2}.
		
		\subsection{The role of separability in the 3-D generalization of model B }\label{sec2.2}
		
		The commutators \eqref{MB-comm1} and  \eqref{MB-comm2} suggest that the operator representing the position random variable in model B, $\hat{X}_t$, resemble the ordinary position operator of non-relativistic quantum mechanics (such operator will be labeled by $\hat{Q}_t$, to avoid ambiguities). In section III E and section IV F of \cite{LC2}, the construction of the Hilbert space representation of the position operator was explained in detail. Since in the 3-D generalization we replace the position random variable with a vector-valued random variable, the Hilbert space on which the operator representing the position random variable acts, is
		\begin{equation*}
		\Hi(X_t) = \Hi(X^{(1)}_t)  \otimes \Hi(X^{(2)}_t)  \otimes \Hi(X^{(3)}_t),
		\end{equation*}
		where $\Hi(X^{(i)}_t)$ is the Hilbert space on which the operator representing the $i$-th component of $X_t$ is defined. Clearly $\Hi(X_t)$ is infinite-dimensional and non-separable in general, because each $\Hi(X^{(i)}_t)$ is not. We recall that on this Hilbert space we can describe also the operator $\hat{V}_t(t')$, but it is not diagonalisable on the same basis of $\hat{X}_t$ (which is diagonal by construction in $\Hi(X_t)$, see section IV F in \cite{LC2}). The importance of the separability was already recognised in section IV G of \cite{LC2} where we showed that, assuming the separability of the infinte dimensional Hilbert space on which the operators are represented, we can relate the velocity random variable of the model to the ordinary momentum operator of non-relaitivistic quantum mechanics $\hat{P}_t$. The same argument can be used also in the 3-D generalisation: assuming the separability of $\Hi(X_t)$, and the existence of a suitable operator $\hat{H}$, by the Ehrenfest theorem we can prove that
		\begin{equation*}
		\lim_{t' \rightarrow t} \hat{V}^{(i)}_t(t') = \frac{1}{m} \hat{P}_t^i
		\end{equation*}
		\emph{in weak sense} for any $i =1,2,3$, where $m$ is a constant with the dimension of a mass, and $\hat{P}_t^i$ is the $i$-th component of $\hat{P}_t$. Summarising, as observed at the end of \cite{LC2}, if we can justify separability then a direct correspondence between the particle in model B and a quantum particle seems to be possible.
		
		\section{Model C: Recovering quantum mechanics} \label{sec3}
		
		In this section, we modify model B in order derive the right commutation relations between position and momentum and establish a correspondence between this new model and non-relativistic quantum mechanics. The model we will obtain will be called model C. Before starting let us point out that the proposed model does not take into account the spin of a quantum particle: this model was not conceived for such a scope. This does not mean that we cannot describe a particle with spin, but simply that the description is done exactly as in ordinary non-relativistic quantum mechanics (hence this model does not add anything to the ordinary description).
		
		The idea behind model C is the following. In this model, we use as space process a DPP, since it is described by using a separable infinite-dimensional Hilbert space. The analysis done in section \ref{sec2}, where we recognize the underlying point process structure of model B, allows to replace the point process used for model B with a DPP in a quite straightforward way. Closure under thinning of the DPP allows to derive the right separable Hilbert space structure for the particle process. Since separability is not enough to establish a correspondence between the velocity operator $\hat{V}_t(t')$ and the quantum momentum operator $\hat{P}_t$, one needs also to introduce a suitable operator $\hat{H}$. This can be done by imposing the correct symmetry group for a non-relativistic system, i.e.  the Galilean group. 
		
		\subsection{The space process}\label{sec3.1}
		
\begin{figure}[h!]
			\includegraphics[height = 230pt,width=250pt]{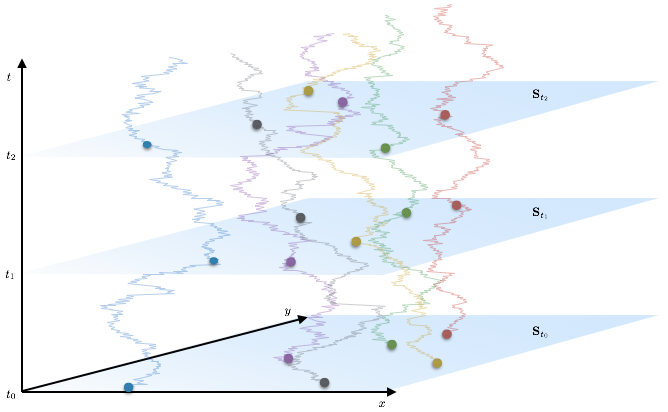}
			\caption{A pictorial representation of a realization of the space process of model C. Note that in model C, the point process at different times is a collection of points in a 3-D space, which here is represented by the 2-D planes in light-blue. The coloured points belonging to the same plane represent a configuration of the DPP at a given time. The trajectories of the DPP diffusion are also drawn in the back.}
			\label{figure1}
		\end{figure}

		We start the description of this model from the space process. At a given time, say $t = 0$, the space process is represented by a DPP on $\Rea^3$ with locally trace-class operator $\hat{K}$ on $L_2(\Rea^3)$. The number of points of the DPP is assumed to be infinite which means, by the theorem \ref{DPP-Noptheo}, that $\mbox{Rank}(\hat{K}) = \infty$. The space process evolves via diffusion, as briefly explained in section \ref{sec1.4}. Such a diffusion preserves the determinantality of the process: this means that if at time $t = 0$ the space process is determinantal it remains so for any $t > 0$, as theorem \ref{DPPdif - fundtheo} shows.
		\begin{definition}
			At a given time $t$ the \emph{space process of model C}, call it $\mathbb{S}^C_t$, is a DPP on $\Rea^3$ with locally trace-class operator $\hat{K}$ on $L_2(\Rea^3)$ having $\mbox{Rank}(\hat{K}) = \infty$. The time evolution of $\mathbb{S}^C_t$, i.e. $t \mapsto \mathbb{S}^C_t$, is a DPP diffusion. The stochastic process $\mathbb{S}^C = \{\mathbb{S}^C_t\}_{t \in \Rea^+}$ is the space process of model C.
		\end{definition}
		A possible realisation of the space process $\mathbb{S}^C$ is given in figure \ref{figure1}.
		
		\subsection{The particle process}\label{sec3.2}
		
	    Let us turn our attention to the particle process at a given time $t$. We have already seen in section \ref{sec2.1} how we can describe the particle process using a marked point process, whose ground process is the point process of the model. Thus we consider the marked point process
	    \begin{equation*}
	    \tilde{\mathbb{S}}_t^C := \{ (x^{(1)}_t , 1), (x^{(2)}_t , 2) , \cdots \},
	    \end{equation*}
	    defined exactly as in section \ref{sec2.1}. According with the discussion done in that section, this can be considered as the physical space equipped with a particular distance function. By \eqref{MPP - CountM},
	    \begin{equation*}
	    N_{\Spc_t}(A) = \sum_{i = 0}^{\infty} N_{\tilde{\mathbb{S}}_t^C}(A \times \{ i \})
	    \end{equation*}
	    which is finite since the ground process is a DPP. This means that, by construction, the sum on the RHS is convergent and so each member of the sum is a point process too. The point process $\{N_{\tilde{\mathbb{S}}_t^C}(A \times \{ i \})\}_{A \in \borell{3}}$ has only a single point and is obtained from the initial DPP, $\Spc_t$, by deleting all the points having mark $j \neq i$. Hence, applying repeatedly theorem \ref{DPP-thin}, we can conclude that $\{N_{\tilde{\mathbb{S}}_t^C}(A \times \{ i \})\}_{A \in \borell{3}}$ is a DPP. Call $\hat{\rho}$ the locally trace-class operator on $L_2(\Rea^3)$ of this DPP. Since it has a single point, we have to conclude that $\mbox{Rank}(\hat{\rho}) = 1$ ( see theorem \ref{DPP-Noptheo}), which means that the kernel of $\hat{\rho}$ can be written as
	    \begin{equation*}
	    \rho(x,y) = \psi(x) \psi^*(y),
	    \end{equation*}
	    where $\psi(x) \in L_2(\Rea^3)$. We introduce a selection random variable, i.e. the positive integer valued random variable $I_t: \Omega_I \rightarrow \Nat$, which allows to define the position of the particle at time $t$ as the point described by the DPP $\xi_{X_t}:=\{N_{\tilde{\mathbb{S}}_t^C}(A \times \{ I_t \})\}_{A \in \borell{3}}.$ 
	    \begin{definition}
	    	Let $\tilde{\mathbb{S}}^C_t$ be the marked point process defined above with ground process $\Spc_t$ and $I_t$ be an $\Nat$-valued random variable. The \emph{position of the particle at time} $t$ in model C is described by the point process on $\Rea^3$ defined as
	    	\begin{equation*}
	    	\xi_{X_t}:= \{N_{X_t}(A)\}_{A \in \borell{3}},
	    	\end{equation*}
	    	where $N_{X_t}(A):=N_{\tilde{\mathbb{S}}_t^C}(A \times \{ I_t \})$ for any $A \in \borell{3}$.
	    \end{definition}
	    Let us label by $X_t$ the (single) point of $\xi_{X_t}$: this is the \emph{position random variable} at time $t$. Again the origin is assumed to be in $\mathbf{0}$, without loss of generality.
	    \begin{definition}
	    	Let $X_t$ and $X_{t'}$ be the position random variables at time $t$ and $t'$ obtained from $\xi_{X_t}$ and $\xi_{X_{t'}}$, respectively. The \emph{velocity random variable} $V_t(t')$ is defined as
	    	\begin{equation*}
	    	V_t(t') := \frac{X_{t'} - X_t}{t' - t}.
	    	\end{equation*}
	    \end{definition}
	    As for model B, $V_t(t')$ should be interpreted as the average velocity of the particle in the time interval $[t,t']$. Also in this case one can introduce the transition probabilities $\alpha(v,x)$. The whole analysis done in section IV B of \cite{LC2} on the relation between the probability distributions of the velocity and position random variables holds also here and we will not repeat it.
	    
	    As already explained, theorem \ref{DPP-thin} implies that $\xi_{X_t}$ is a DPP with a rank-1 kernel. This means that the intensity measure of this process, is
	    \begin{equation*}
	    M_1(A) = \int_A \rho(x,x)dx = \int_A |\psi(x)|^2 dx.
	    \end{equation*}
	    Since $\mu_{X_t}(A) := M_1(A)$ is the expected number of times one finds the point of $\xi_t$ in $A$, we can conclude that $X_t$ has probability density $|\psi(x)|^2$ for some $\psi(x) \in L_2(\Rea^3)$. However, as in model A and model B, we are not interested in the probability $P[X_t \in A] = \mu_{X_t}(A)$, but rather in the conditional probability with respect to some configuration of the space process. Let $\sss_t$ label a possible configuration of the marked point process $\tilde{\mathbb{S}}_t^C$ with fidis density $\mu(\sss_t)$, namely we can write
	    \begin{equation*}
	    \mu(A_1,n_1;\cdots;A_r,n_r) = \int_{A_1^{(n_1)}\times \cdots \times A_r^{(n_r)}} \mu(\sss_t)d\sss_t.
	    \end{equation*}
	    As explained in section \ref{sec1.1}, the fidis can be expressed in terms of the Janossy measure, thus the fidis density expressed above can be obtained from the Janossy densities (which always exist for a DPP). We note that once $\sss_t$ is fixed, the position random variable can vary only over a discrete set of points, i.e. over the set $\Lambda_{\sss_t} := \{x | x \in \sss_t\}$ which is a subset of $\Rea^3$. Thus we can write
	    \begin{equation*}
	    \mu_{X_t}(A) = \int_A |\psi(x)|^2 dx = \int_{\mathcal{N}_{\Rea^3}}\int_{A_{\sss_t}} |\psi(y,\sss_t)|^2 d\sss_t dy,
	    \end{equation*}
	    where $A_{\sss_t}$ is the subset of $\Lambda_{\sss_t}$ compatible with the event $\{X_t \in A\}$. As a matter of fact the integral with respect to $y$ is actually a sum of discrete values. As already observed in the conclusion of \cite{LC2}, in order to deal with probability measures absolute continuous with respect to Lebesgue we need to take the \emph{dense-point limit}. In this limit, the point of space gets denser and denser till the continuum  ($\Rea^3$) is reached. Let us describe more formally how this limit may be taken. The (local) density of the space process $\Spc_t$, can be measured by counting the number of points in an arbitrary set $A \subset \Rea^3$, namely with $n:= N_{\Spc_t}(A)$. The dense point limit is obtained for $n \rightarrow \infty$ for any $A \subset \Rea^3$ (when $n = \infty$ for any $A \subset \Rea^3$ the point process structure of the space is replaced by the continuum). The density $n$ parametrises also the marked point process $\tilde{\mathbb{S}}_t^C$ (we write $\tilde{\mathbb{S}}_t^C[n]$). The function $y \mapsto \psi(y,\sss_t)$ is in $L_2(\Lambda_{\sss_t[n]})$ for any $n$ finite. Since $n = \infty$ corresponds to the continuum  $\Rea^3$ and $L_2(A) \subset L_2(B)$ if $A \subset B$ \footnote{We are assuming that a function $f(x)$ in $L_2(A)$ is extended to $L_2(B)$ setting $f(x) = 0$ when $x \in B/A$.}, in the dense point limit we have $y \mapsto \psi(y,\sss_t) \in L_2(\Rea^3)$. From now on, we assume to work in the dense-point limit of the space process. In this case we write
	    \begin{equation*}
	    \begin{split}
	    \mu_{X_t}(A) &= \int_{\mathcal{N}_{\Rea^3}}\int_{A} |\psi(y,\sss_t)|^2 d\sss_t dy \\
	    &= \int_{\mathcal{N}_{\Rea^3}}\int_{A} |\psi_{\sss_t}(y)|^2  \mu(\sss_t)d\sss_tdy
	    \end{split}
	    \end{equation*}
	    where
	    \begin{equation}\label{wavefunction}
	    \psi_{\sss_t}(y) := \frac{\psi(y,\sss_t)}{\sqrt{\mu(\sss_t)}}.
	    \end{equation}
	    The conditional probability density $\mu_{X_t|\sss_t}(x) = |\psi_{\sss_t}(x)|^2$ is the probability measure that we can use, repeating the arguments used for model B (see section IV E and section IV F in \cite{LC2}), to represent the random variable $X_t$ with an operator on an Hilbert space $\Hi(X_t)$. After conditioning on $\sss_t$, since $\psi(x,\sss_t) \in L_2(\Rea^3)$ as function of $x$, then also $\psi_{\sss_t}(x)$ is a square integrable function on $\Rea^3$, i.e. $\psi_{\sss_t}(x) \in L_2(\Rea^3)|_{\sss_t}$\footnote{It is worth to note that $\psi_{\sss_t}(x)$ is a square integrable function on $\Rea^3$, thus is an element of $L_2(\Rea^3)$. However, to emphasise the dependence on the configuration of the DPP and distinguish such Hilbert space from the $L_2(\Rea^3)$ space of the DPP, we choose to use the symbol $L_2(\Rea^3)|_{\sss_t}$. }. We have that
	    \begin{equation}\label{Ex-Vop}
	    \begin{split}
	    \Ex[X_t | \sss_t] &= \int_{\Rea^3} x |\psi_{\sss_t}(x)|^2 dx \\
	    &= \int_{\Rea^3} \psi_{\sss_t}^*(x) x \psi_{\sss_t}(x) dx \\
	    &= \langle \psi_{\sss_t} | \hat{X}_t \psi_{\sss_t} \rangle,
	    \end{split}
	    \end{equation}
	    where $\hat{X}_t$ is a self-adjoint multiplicative operator on $L_2(\Rea^3)|_{\sss_t}$ whose action is
	    \begin{equation}\label{Xop}
	    \hat{X}_t \psi_{\sss_t}(x) = x \psi_{\sss_t}(x),
	    \end{equation}
	    and with domain given by
	    \begin{equation*}
	    D(\hat{X}_t) := \{ \phi \in L_2(\Rea^3)|_{\sss_t} \st \hat{X}_t \phi \in L_2(\Rea^3)|_{\sss_t} \}.
	    \end{equation*}
	    The domain is defined such that $\hat{X}_t$ is a well defined operator on $L_2(\Rea^3)|_{\sss_t}$. This is exactly the operator representing $X_t$ on an Hilbert space, when we follow the procedure explained in \cite{LC} and \cite{LC2}. Thus we can conclude that the DPP process induces the separability of the Hilbert space $\Hi(X_t)$ of model C. It is not difficult to note that $\hat{X}_t$ coincides with the ordinary position operator $\hat{Q}_t$ in non-relativistic quantum mechanics. Finally, also the velocity random variable can be represented on the same Hilbert space, i.e. $\hat{V}_t(t')$ is an operator acting on $L_2(\Rea^3)|_{\sss_t}$, as in the case of model B. Separability ensures that starting from the velocity random variable and constructing the Hilbert space $\Hi(V_t)$, the operators representing the particle process on this Hilbert space coincide with the position and velocity operators defined on $\Hi(X_t)$. Summarising, on $L_2(\Rea^3)|_{\sss_t}$ the two operators associated to the particle process are defined. In particular, the position of the particle process coincides with the quantum mechanical operator. Nothing, for the moment, can be said about the relation between $\hat{V}_t(t')$ and the momentum operator of non-relativistic quantum mechanics.
		
		\subsection{Recovering the momentum operator}\label{sec3.3}
		
		In the previous section, we showed that when the space process is a DPP, the position and velocity random variables, after conditioning, are represented by operators on a separable infinite-dimensional Hilbert space. In section IV G of \cite{LC2}, we argue that separability, together with the existence of a suitable operator $\hat{H}$, implies that for $t' \rightarrow t$, the operator representing the velocity random variable, reduces to the momentum operator of ordinary non-relativistic quantum mechanics. Here we repeat the argument in a more systematic way, and we introduce the required $\hat{H}$ operator via symmetry group considerations.
		
		We want to show that the operator representing the velocity random variable $\hat{V}_t(t')$, reduces to the ordinary momentum operator of non-relativistic quantum mechanics $\hat{P}_t$. In order to do that, we can use the Ehrenfest theorem. We will not state it in its whole generality, but we consider only the part of this theorem regarding the position operator in quantum mechanics (see Cor 1.2 and Th 1.3 \cite{friesecke2009ehrenfest}).
		\begin{theorem}\label{Ehre-theorem}
			Let $\hat{H}$ be an operator on an Hilbert space $L_2(\Rea^d)$ of the from
			\begin{equation*}
			\hat{H} = - \sum_{i = 1}^d \frac{1}{2 m_i} \frac{\partial^2}{\partial q_i^2} + V(q)
			\end{equation*}
			with domain $D(\hat{H}) = H_2(\Rea^d)$ \footnote{$H_2(\Rea^d)$ is the Sobolev space of square integrable functions on $\Rea^d$. }
			where $m_1,\cdots,m_d > 0$, $V: \Rea^d \rightarrow \Rea$ is a real valued, locally integrable function fulfilling the Kato-Rellich condition\footnote{Let $\hat{G}: D(\hat{G}) \rightarrow \Hi$ and $\hat{V}: D(\hat{V})\rightarrow\Hi$ be densely defined operators on the Hilbert space $\Hi$. If there exist $a,b \in [0,\infty)]$ such that
				\begin{equation*}
				\|\hat{V}\varphi\| \leqslant a \| \hat{G} \varphi \| + b \| \varphi \|,
				\end{equation*}
				for any $\varphi \in D(\hat{G})$, then $\hat{V}$ is said $\hat{G}$\emph{-bounded}. The greatest lower bound of all the numbers $a$ satisfying the condition above for some $b$, is said \emph{relative bound} of $\hat{V}$ with respect to $\hat{G}$. We say that $\hat{V}$ satisfies the \emph{Kato-Rellic condition} if, given
				\begin{equation*}
				\hat{G} = - \sum_{i = 1}^d \frac{1}{2 m_i} \frac{\partial^2}{\partial q_i^2},
				\end{equation*}
				the $\hat{V}$ is $\hat{G}$-bounded with relative bound $a < 1$.}. Let $\psi_t$ be a solution of the Schr\"{o}dinger equation having hamiltonian $\hat{H}$ written in the $\hat{Q}^i$-representation, where $\hat{Q}^i$ the $i$-th component of the position operator in non-relativistic quantum mechanics. Then $\langle \hat{Q}^i\rangle_{\psi_t}$ is continuously differentiable with respect to $t$ for any $\psi_t \in D(\hat{H}) \cap D(\hat{Q}^i_t)$ and satisfies the equation
			\begin{equation*}
			\frac{d}{dt} \langle \hat{Q}^i \rangle_{\psi_t} = i \left( \langle \hat{H} \psi_t | \hat{Q}^i \psi_t \rangle - \langle \hat{Q}^i \psi_t | \hat{H} \psi_t \rangle \right).
			\end{equation*}
			Moreover, if $V:\Rea^d \rightarrow \Rea$ has also locally weak derivative which is integrable, and also $\sqrt{|\nabla V|}$ satisfies the Kato-Rellich condition, the equation above reduces to
			\begin{equation}\label{Eheren-theo}
			\frac{d}{dt} \langle \hat{Q}^i \rangle_{\psi_t} =\frac{1}{m_i} \langle \hat{P}^i \rangle_{\psi_t}.
			\end{equation}
		\end{theorem}
		First we observe that in model C, $\mu_{X_t|\sss_t}(x) = |\psi_{\sss_t}(x)|^2$ depends on time (it represents the probability distribution of the position of the particle \emph{at time $t$}), thus one should write $\psi_{\sss_t}(x,t)$ instead of $\psi_{\sss_t}(x)$: such time dependence was omitted in section \ref{sec3.2}. However, the operator $\hat{X}_t$ associated to the random variable $X_t$ always fulfils \eqref{Xop}, for any time $t$. In this sense, it does not evolve in time. Now, computing the expectation value of the velocity random variable, we can write
		\begin{equation*}
		\begin{split}
		\Ex[V_t(t')|\sss_t]& = \frac{\Ex[X_{t'} | \sss_{t}] - \Ex[X_t | \sss_t]}{t' - t} \\
		& = \frac{\langle \hat{X}_{t'} \rangle_{\psi_{t'}} - \langle \hat{X}_t \rangle_{\psi_t}}{t' - t},
		\end{split}
		\end{equation*}
		where we used \eqref{Ex-Vop} and we omitted the $\sss_t$ label in the last line. As already observed at the end of \cite{LC2}, the expectation values used in the first line of the above formula are defined on different probability spaces: the three random variables cannot be defined on the same probability space after the space process ($\tilde{\mathbb{S}}_t^C$ in this case) is removed. Taking the limi $t' \rightarrow t$, we get
		\begin{equation*}
		\lim_{t' \rightarrow t} \Ex[V_t(t')|\sss_t] = \lim_{t' \rightarrow t} \frac{\langle \hat{X}_{t'} \rangle_{\psi_{t'}} - \langle \hat{X}_t \rangle_{\psi_t}}{t' - t} = \frac{d}{dt}\langle \hat{X}_t \rangle_{\psi_t}.
		\end{equation*}
		Since $\hat{Q}_t = \hat{X}_t$, equation \eqref{Eheren-theo} suggests that
		\begin{equation*}
		\lim_{t' \rightarrow t} \Ex[V_t(t')|\sss_t] = \lim_{t' \rightarrow t} \langle \hat{V}_t(t') \rangle_{\psi_t} = \frac{1}{m} \langle \hat{P} \rangle_{\psi_t}.
		\end{equation*}
		At the level of operators, if the relation above holds, we can say that $\lim_{t' \rightarrow t} \hat{V}_t(t')  = \hat{P}_t / m$ \emph{weakly}. In order to state that, we need to introduce an operator $\hat{H}$ on $L_2(\Rea^3)|_{\sss_t}$ fulfilling the conditions in theorem \ref{Ehre-theorem}.
		
		To introduce $\hat{H}$ in a physically meaningful manner, we can proceed using the symmetry group. In particular, since the symmetry group of non-relativistic quantum mechanics is the (central extension) of the Galilean group (see \cite{bargmann1954unitary}, or Ch. 12 of \cite{moretti2013spectral}), we may impose that the probabilities in our model are invariant under this symmetry group. A subgroup of the Galilean group is the abelian additive group $(\Rea, +)$ of time translations. The operator $\hat{H}$ can be introduced, via Stone's theorem, as the generator of the time translation of the Galilean group. In particular, for a single particle, it is
		\begin{equation}\label{Galilean hamiltonian}
		\hat{H} = \frac{1}{2m} \hat{P}^2
		\end{equation}
		where $\hat{P}$ is the momentum operator in non-relativistic quantum mechanics and such operator fulfils the requirements of theorem \ref{Ehre-theorem}. This is nothing but the kinetic term for a non-relativistic free quantum particle. This means that, if we demand that the probabilistic interpretation of the model remains invariant under the Galilean subgroup of time translations, then $\psi_{\sss_t}(x,t) \in L_2(\Rea^3)|_{\sss_t}$ evolves in time as
		\begin{equation*}
		\psi_{\sss_{t'}}(x,t') = e^{-it\hat{H}} \psi_{\sss_t}(x,t),
		\end{equation*}
		or equivalently
		\begin{equation*}
		i \frac{d}{dt} \psi_{\sss_t}(x,t) = \hat{H} \psi_{\sss_t}(x,t),
		\end{equation*}
		which is the usual Schr\"{o}dinger equation for a free non-relativistic quantum particle. Clearly this time evolution for the probability of $X_t$ is not a consequence of model C only, but is a general feature of the Galilean group. In addition by, representing the Galilean group on $L_2(\Rea^3)|_{\sss_t}$,we also obtain the operator representation of other physical observables like, for example, the angular momentum. Summarising, by using considerations on the symmetry group of non-relativistic systems one can introduce the operator $\hat{H}$ with the correct properties in a physically meaningful way. This justifyes the conclusion that for a non-relativistic particle, 
		\begin{equation*}
		\lim_{t' \rightarrow t} \hat{V}_t(t')  = \frac{\hat{P}_t}{m}
		\end{equation*}
		\emph{weakly}, for some constant $m > 0$ having the dimension of a mass. In this way, we may recover the quantum mechanical momentum operator in model C. Note that this is not a consequence of the Ehenefest theorem and Galilean group only: model C is capable to reproduce the correct Hilbert space structure, \emph{without postulating it in advance}. Because of this we can conclude that, model C \emph{together with the Galilean group} is capable to reproduce the non-relativistic quantum mechanics of a free particle.
		
		\subsection{$n$-particle generalization}\label{sec3.4}
		
		In the previous section, we arrived at the conclusion that we can use model C to describe a free single quantum particle. Here we want to extend the discussion to the $n$-particle case. In particular, we want to understand if we can derive from model C the \emph{tensor product postulate} of non-relativistic quantum mechanics, namely the fact that the Hilbert space associated to two distinguishable particles is the tensor product of the single-particle Hilbert spaces. Here we will consider the $2$-particle case only, since considerations to the case $n > 2$ can be straightforwardly generalized. As we will see, despite  model C is capable to describe two or more particles, we cannot fully derive the tensor product postulate without adding further conditions (or at least the author was not able to do so).
		
		Following the idea explained in section \ref{sec3.2}, where we discussed the position random variable for a single particle, we generalize the description to the case of two particles in the following natural way. Given the space process $\Spc_t$ on $\Rea^3$ at a given time $t$, let us consider two marked DPPs $\tilde{\mathbb{S}}^C_t(a)$ and $\tilde{\mathbb{S}}^C_t(b)$, one for each particle, having both as ground process $\Spc_t$. Then we consider the point process obtained from the cartesian product of $\tilde{\mathbb{S}}^C_t(a)$ and $\tilde{\mathbb{S}}^C_t(b)$, namely
		\begin{equation*}
		\overline{\mathbb{S}}^C_t := \{ (x_t^{(1)},1) \times (y_t^{(1)},1) , (x_t^{(1)},1) \times (y_t^{(2)},2) , \cdots \}
		\end{equation*}
		where $x_t^{(i)}$ and $y_t^{(j)}$ are points of $\Spc_t$. The ground process of $\overline{\mathbb{S}}^C_t$ has counting measure
		\begin{equation*}
		N_g(A \times B) = \sum_{i = 1}^\infty \sum_{j = 1}^\infty N_{\overline{\mathbb{S}}^C_t }(A \times \{i\} \times B \times \{j\}),
		\end{equation*}
		which is finite, because $N_g(A \times B)$ is just the number of points of $\Spc_t \times \Spc_t$ in $A \times B$, and $\Spc_t$ is a DPP. Thus each component must be $N_{\overline{\mathbb{S}}^C_t }(A \times \{i\} \times B \times \{j\}) < \infty$, and so it defines a point process. Such process can be obtained from the original point process $\Spc_t$ by removing all the points having labelling different from $i$ and $j$. By theorem \ref{DPP-thin}, the thinned point process is again a DPP and has \emph{at least two points}. However, we can always say that the point process has just a \emph{single point} if we consider, as ground process of $\overline{\mathbb{S}}^C_t$, the cartesian product of two copies of $\Spc_t$. This is the same observation done in section \ref{sec1.1} which allows to interpret the moment measure $M_2$ for a point process $\xi$ as the intensity measure $M_1$ of the point process $\xi \times \xi$. This means that for $i$ and $j$ fixed, the counting measure $N_{\overline{\mathbb{S}}^C_t }(A \times \{i\} \times B \times \{j\})$ can define a point process on $\Rea^3 \times \Rea^3 = \Rea^6$ having a single point. Now, we introduce two selection random variables, $I_t : \Omega \rightarrow \Nat$ and $J_t: \Omega \rightarrow \Nat$, which may depend on each other, and we define the point process on $\Rea^6$
		\begin{equation*}
		\xi_{(X_t^{a},X_t^{b})} := \{N_{\overline{\mathbb{S}}^C_t }(A \times \{I_t\} \times B \times \{J_t\})\}_{A\times B \in \borell{6}},
		\end{equation*}
		as done in section \ref{sec3.2}. The joint probability distribution for the positions of two particles at a given time $t$ will be
		\begin{equation*}
		\mu_{(X_t^{a},X_t^{b})}(A \times B) = \int_{A \times B} \rho_1(x,y)dxdy,
		\end{equation*}
		where $\rho_1(x,y)$ is the intensity measure density of the point process $\xi_{(X_t^{a},X_t^{b})} $. At this point a difficulty arises: we are not able to deduce the that $\rho_1(x,y) = |\psi(x,y)|^2$ with $\psi(x,y) \in L_2(\Rea^6) = L_2(\Rea^3) \otimes L_2(\Rea^3)$. The reason for this difficulty lies in the fact that, if  a point process $\xi$ is a DPP, then the point process $\xi \times \xi$ is not in general a DPP. As consequence, we cannot define the Hilbert space $L_2(\Rea^6)|_{\xi_t}$, and so in general the Hilbert space on which the operators associated to the two position random variables are defined, may be non-separable. More precisely, we can have two cases:
		\begin{enumerate}
			\item[a)] The point process $\overline{\mathbb{S}}^C_t$ is determinantal. If so, as in the single particle case, we know that there exist $\psi(x,y) \in L_2(\Rea^6) = L_2(\Rea^3) \otimes L_2(\Rea^3)$ such that
			\begin{equation*}
			\mu_{(X_t^{a},X_t^{b})}(A \times B) = \int_A \int_B |\psi(x,y)|^2 dxdy.
			\end{equation*}
			In this case, using \eqref{wavefunction}, we obtain that the (conditional on $\sss_t$) probability density of the position of the two particles, $\mu_{(X_t^{a},X_t^{b}) | \sss_t}$, is given by the square modulus of a function in $L_2(\Rea^3) \otimes L_2(\Rea^3)$. At this point depending on whether $\mu_{(X_t^{a},X_t^{b}) | \sss_t}$ factorise, i.e.  $\mu_{(X_t^{a},X_t^{b}) | \sss_t} = \mu_{X_t^{a} | \sss_t}\mu_{X_t^{b} | \sss_t}$, or not, one can describe particles which are uncorrelated or entangled. 
			\item[b)] The point process $\overline{\mathbb{S}}^C_t$ is not determinantal. In this case we are in a situation which is similar to the one of model B. Once we choose to represent the position random variables with operators, the Hilbert space on which both these operators acts, $\Hi(X_t^{a},X_t^{b})$, is in general non-separable, which implies that $\Hi(X_t^{a},X_t^{b}) \supset L_2(\Rea^3) \otimes L_2(\Rea^3)$. However, also in this case we can describe uncorrelated or entangled-like particles depending on whether the probability distribution $\mu_{(X_t^{a},X_t^{b})}(A \times B)$ after conditioning on $\sss_t$ (which, in any case, defines a state on the algebra describing two particles in model C) factorise or not.
		\end{enumerate}
		We have to admit that the case $a)$ seems to be very rare. Thus, we conclude that using the natural generalization of the procedure showed in section \ref{sec3.2}, \emph{we cannot say that the Hilbert space on which we can describe two particles is always the tensor product of the single-particle Hilbert spaces}. To overcome this difficulty we suggest the following approach. Instead of deriving the tensor product structure from the point process structure of space, one may solve this difficulty \textquotedblleft probabilistically", as explained below. For a single particle, the ordinary quantum mechanical description of probability applies. This means that the whole machinery of quantum logic and its propositional calculus can be used to describe a single particle in model C. This is almost sufficient to derive the tensor product postulate. In fact, following the work of D. Aerts and I. Daubechies \cite{aerts1978physical}, \emph{within the quantum logic framework}, given two quantum systems and requiring that 1) the structure of the two systems is preserved; 2) a measurement on one system does not disturb the other (in the sense that operators associated to the two systems commute); 3) maximal information obtained on both system separately gives the maximal information on the joint system; one can conclude that the Hilbert space of the joint quantum system is the tensor product of the single particle Hilbert space. Summarizing, by adding these assumptions, the tensor product postulate of quantum mechanics can be used within the framework of model C.

		\subsection{Does the Bell and PBR no-go theorems apply to model C?}\label{sec3.5}
		
		The work done till now shows that model C can be put in correspondence with ordinary non-relativistic quantum mechanics. However, the models proposed in \cite{LC2} and model C use essentially classical probability theory (more precisely, the Kolmogorov's axiom of probability theory), a possibility which seems to be excluded by various no-go theorems. Here we will argue why  Bell's theorem \cite{bell1964einstein} and PBR theorem\cite{pusey2012reality} do not rule out the models proposed and in particular model C. 
		
		In \cite{LC} we described the general probabilistic framework in which the three models were elaborated. In particular we introduce the notion \emph{contextual probabilistic model} elaborated in \cite{khrennikov2009contextual} and \cite{khrennikov2016probability}. In \cite{LC2} and also here, each random variable is considered on its own probability space. The position $X_t$ and the velocity $V_t(t')$ are defined initially on a common probability space however, we describe them only when the space is removed. This changes the probability space on which these random variables are considered. $X_t$ is defined on a probability space obtained by conditioning on $\sss_t$. Also the velocity $V_t(t')$ can be defined on a probability space obtained conditioning on $\sss_t$, but is not the same of $X_t$ because the Bayes theorem does not hold (we use the unconditional transition probabilities). Thus each random variable is considered on its own probability space: this is a consequence of the removal of the space process. The entropic uncertainty relations that we found in the models enforce exactly this conclusion: the position and the velocity random variable \emph{must be in two different probability spaces}. Both Bell's and  PBR theorem assume that all the observables can be represented as random variables acting on the \emph{same} probability space (where the Bayes theorem holds): this is the reason why the two theorems cannot be applied directly. For the analysis of the probabilistic arguments used in the Bell's theorem we refer to \cite{khrennikov2009contextual, khrennikov2016probability, khrennikov2014bell}. There it is explained how fundamental is the assumption of a single probability space for the validity of the Bell's argument. A direct application of the PBR argument to model C is not possible at the moment. However since in model C each observable is in its own probability space, the PBR theorem does not seem to be applicable in this case.
		
		\section{Summary and critics}\label{sec4}
		
	    In this section we list and criticize some of the assumptions used in the models presented in \cite{LC2} and in model C.  Let us start with an informal summary of the work done. In \cite{LC2}, we declare our intention to explore the possibility that the commutation relations $[\hat{Q}^i,\hat{P}^i] = i \Id$, are not due to an irreducible disturbance generated by the interaction of the quantum system with the measurement apparatus, but rather are consequence of an intrinsic stochasticity of the physical space on which the particle exists. Possible arguments to support this idea were given by using the quantum ruler, i.e. an attempt to describe at the quantum level a ruler. Assumed that, we started with the description of the space which was thought as a random distribution of points in a given \emph{mathematical} space ($\Int$, $\Rea$ or $\Rea^3$). The physical space was always assumed to be the random distribution of points and not the underlying mathematical space. On the physical space, a particle was defined through two random variables: one for its position and one for its velocity at a given time (which is always an external parameter). This particle moves by jumps from one point of the physical space to the other. Once the physical space is not taken into account in the probabilistic description of the particle (operation that was implemented by conditioning on \emph{a} space configuration, and not averaging over \emph{all} the space configurations), the mathematical model that we can to use to describe the particle is a non-commutative probability theory (we can no longer use the ordinary measure-theoretic description consisting in a single probability space). Imposing conditions on the space process (requiring to be a DPP) and on the group of symmetry under which the particle is invariant (the symmetry group is the Galilean one), we showed that the description of the particle in this framework coincides with that of a non-relativistic quantum particle. Below we list and discuss some possible critics on the proposed model.
	    \begin{enumerate}
	    	\item[i)] \emph{Space process.} We chose to model space with a random distribution of points over a $\Rea^3$. Nothing forbid to start from some kind of \textquotedblleft \emph{stochastic manifold}", in the sense of a manifold on which the distance between points changes at random. For an interesting work in this direction, one may consult \cite{frederick1976stochastic,roy2012statistical}. Other interesting structures can be \emph{graphs}, namely a collection of points (called \emph{vertices}) and relations between two points which can be visualized with a straight line linking the two related points (called \emph{edge}). To implement the intrinsic randomness of space, one can imagine to have a probability distribution over the set of all edges, describing the probability that an edge, linking point $a$ to point $b$, exists. This last approach is particularly interesting since it leaves open the possibility to recover the underlying mathematical space $\Rea^3$ with some limiting procedure \cite{bombelli2009semiclassical}. We also note the following curious fact: the dimensionality of the problem is dictated by the dimensionality of the underlying mathematical space and not of the physical space. The use of graphs for the description of the space process may also help to remove this unpleasant feature of the model, for example, using techniques similar to those used for causal sets \cite{bombelli1987space,brightwell1991structure,reid2003manifold}.
	    	\item[ii)] \emph{Hilbert space representation.} As already discussed in \cite{LC} and \cite{LC2}, the Hilbert space representation of the random variables of the model is obtained via algebric methods. Only in some cases, we can represent \emph{explicitly} the probability distribution with vectors of the Hilbert space, i.e. in all the cases where QRLA applies. However, only for dichotomic random variables, QRLA is fully developed, while for trichotomic random variables QRLA is only partially developed. In \cite{aerts2001probability}  an interesting possibility of how to represent a probability distribution with vectors in $L_2(\Rea^3)$ (hence a possible starting point to generalise QRLA to continuous random variables) is presented: there a vector in $L_2(\Rea^3)$ from a probability distribution is constructed but nothing is said about a change of basis. This means that, in principle, given a probability distribution of the position, we can construct $\psi(x)$, but we do not know how to change basis and derive the probability distribution of the velocity. In the limit where the velocity of the model coincides with the quantum mechanical momentum, we know how a change of basis can be performed, but in the general case, the algorithm is not available. 
	    	\item[iii)] \emph{Determinantal point process.} In order to induce the correct Hilbert space structure, i.e. an infinite dimensional separable Hilbert space, we used a specific point process: the DPP. Thanks to this structure we were able to derive the correct description of a free quantum particle. However, since space is \textquotedblleft removed" from the models, it seems strange that space still plays such a fundamental role. For example, there are other point processes, like the permanental or more generally $\alpha$-determinantal point process \cite{fukushima2010dirichlet}, which use kernels of locally trace-class operators in $L_2(\Rea^3)$. Thus one may expect to obtain the quantum description of a single particle also using these processes. However, results which hold for a DPP do not hold anymore for these processes and so technical difficulties may arise. Moreover, also the necessity of special classes of point processes to induce the $L_2$ structure seems to be a rather artificial condition: probably solving the Hilbert space representation problem explained in $ii)$ or some algebraic consideration, may remove this constraint leaving us free to choose any point process we like.
	    	\item[iv)] \emph{Tensor product postulate.}  As observed in section \ref{sec3.4}, model C has difficulties in justifying the tensor product postulate. This is not surprising since we use DPP, to induce the correct Hilbert space structure: when we cannot use anymore the DPP, the problem of the correct Hilbert space structure appears. The way out proposed uses results of quantum logic to justify such a postulate \cite{aerts1978physical}. Because quantum logic can be considered as the background logic of probability theory describing a quantum system, this solution has a \textquotedblleft probabilistic flavour". Note that the necessity to appeal to a more detailed analysis of the probabilistic structure of the models, is in agreement with the observations done in $ii)$ and $iii)$ on the $L_2$ structure.
	    	\item[v)] \emph{Dependence of the Hilbert space on the space configuration.} When we constructed the Hilbert space associated to a single particle in section \ref{sec3.2}, we obtained a $\sss_t$-dependent object. If such dependence is relevant for the evolution of the state vector associated to the probabilities of the model more general time evolution of $\psi_{\sss_t}$ are possible, for example a stochastic unitary time-evolution.
	    \end{enumerate}
		
		\section{Conclusion and future outlooks}\label{sec5}
		
		The critical points recognized in the previous sections are of two kinds: geometric and probabilistic. From the probabilistic side, a general version of QRLA (which can be applied to any couple of random variables) may solve the criticality $ii)$, $iii)$ and $v)$. On the other hand, the geometric problem mentioned in $i)$ strays in the active research field of quantum gravity. The models proposed are interesting because despite space is removed to get the quantum description, it is an active player of the whole model. Space possesses its own dynamics, something which resembles the situation we have in general relativity (despite the space dynamics used here was not chosen to mimic any general relativistic effect).  Moreover, starting from a manifold equipped with a stochastic metric, the stochastic mechanics \cite{nagasawa2012schrodinger,nagasawa2012stochastic,nelson1966derivation} can be obtained (see \cite{roy2012statistical} for an overview). It is interesting to observe that a similar picture, of a particle jumping at random on the space, emerge in stochastic mechanics from a careful analysis of the stochastic trajectories of a quantum particle\cite{nagasawa2012stochastic,nagasawa1989transformations}. 
		To conclude, we think that an interesting part of the new description presented here is the model of space used. Maybe, by applying this new ingredient in contests which are outside the realm of quantum mechanics, can give rise to new interesting results. For example, this idea of space may be used in the contest of the early universe, where the effects of such structure of space may be relevant.
		
		\section{Acknowledgements}
		
		The author is indebted with S. Bacchi for its patience during the discussions about the Galilean group and many other random questions of mathematical nature. Indirect contributions to this work come from  L. Andreoli, A. Endrizzi, E. De Antoni, L. Festa, C. Hille, P. Mori and G. Navarra.
		
		\section{Appendix}
		
		\subsection*{A. Integral kernels, locally trace-class operators and Fredholm's determinants }
		
		This appendix is based on the review \cite{karambal2010introductory} and on Ch. 4 of \cite{moretti2013spectral}. A concise explanation of the topics presented here can also be found in \cite{decreusefond2016determinantal}. Let $\Hi$ be an Hilbert space and $\langle \cdot | \cdot \rangle$ denote its scalar product. As usual $\boundo$ is the set of all bounded operators on $\Hi$, while the set of all compact operators on $\Hi$ will be denoted by $\compo$ and we recall that $\compo \subset \boundo$.
		\begin{definition}
			Let $\Hi$ be an Hilbert space (possibly infinite dimensional and non-separable) and $\|\cdot\|$ denote the norm induced by its scalar product. An operator $\hat{A} \in \boundo$ is said \emph{Hilbert-Schmidt operator} if there is a basis $\{ u_i \}_{i \in I}$ such that
			\begin{equation*}
			\sum_{i \in I} \|\hat{A} u_i\|^2 < \infty.
			\end{equation*}
		\end{definition}
		The set of all Hilbert-Schmidt operators on $\Hi$ will be labeled by $\hso$. One can prove that $\hso \subset \compo \subset \boundo$. On $\hso$ a norm can be defined
		\begin{equation*}
		\| \hat{A} \|_2 := \sqrt{\sum_{i \in I} \|\hat{A} u_i\|^2 },
		\end{equation*}
		and is called \emph{Hilbert-Schmidt norm}. It does not depend on the chosen basis $\{ u_i  \}_{i \in I}$. Such norm makes $\hso$ a Banach space. More generally, on $\hso$ one can show that 
		\begin{equation*}
		\langle \hat{A} , \hat{B} \rangle_2 := \sqrt{\sum_{i \in I} \langle \hat{A}u_i | \hat{B}u_i \rangle}
		\end{equation*} 
		is an inner product. It induces the Hilbert-Schmidt norm and so $\hso$ is an Hilbert space when equipped with such inner product. Hilbert-Schmidt operators are important because of the following theorem.
		\begin{theorem}\label{apA:Kernel-theorem}
			If $\mu$ is a positive $\sigma$-additive separable measure over a locally compact second countable Hausdorff space $\X$, the space $\mathcal{B}_2(L_2(\X, \mu))$ consists of the operators $\hat{A}_K$ defined as
			\begin{equation*}
			(\hat{A}_K f)(x) := \int_{\X} K(x,y) f(y) \mu(dy),
			\end{equation*}
			for any $f(x) \in L_2(\X,\mu)$, where $K:\X \times \X \rightarrow \Comp$ is in $L_2(\X \times \X, \mu \otimes \mu)$. Moreover
			\begin{equation*}
			\| \hat{A}_K \|_2 = \| K \|_{L_2(\X\times \X, \mu \otimes \mu)}.
			\end{equation*}
			In particular, if $\hat{A} \in \mathcal{B}_2(L_2(\X, \mu))$ and $\{u_i \}_{i \in I}$ is a basis of $L_2(\X ,\mu)$, then $\hat{A} = \hat{A}_K$ for
			\begin{equation}\label{apA:kernel-eq}
			K(x,y) = \sum_{i,j \in I} \langle u_i(x)| \hat{A} u_j(y) \rangle u_i^*(x) u_j(y),
			\end{equation}
			and the convergence is in $L_2(\X \times \X, \mu \times\mu)$. The map $K \mapsto \hat{A}_K$ is an isomorphism between the Hilbert spaces $L_2(\X \times \X, \mu \times \mu)$ and $\mathcal{B}_2(L_2(\X, \mu))$.
		\end{theorem}
		From now on, we assume that $\Hi = L_2(\X,\mu)$ (hence it is separable), where $\X$ and $\mu$ are as in theorem \ref{apA:Kernel-theorem}. Since if $\hat{A} \in \mathcal{B}_2(\Hi)$, we can write it as
		\begin{equation*}
		\hat{A} f(x) = \int_{\X} K(x,y)f(y) \mu(dy).
		\end{equation*}
		Hilbert-Schmidt operators are also called \emph{integral operators}. The function $K(x,y): \X \times \X \rightarrow \Comp$ defined as in theorem \ref{apA:Kernel-theorem} is said \emph{Kernel of the operator $\hat{A}$}.
		\begin{proposition}
			If $\hat{A} \in \hso$ is self-adjoint, then the associated kernel $K: \X \times \X \rightarrow \Comp$ satisfies
			\begin{equation*}
			K(x,y) = K(y,x)^*
			\end{equation*}
			for any $x,y \in \X$.
		\end{proposition}
		When we deal with the space $\X$ which is also compact, a useful theorem is the \emph{Mercer theorem}, which simplify further equation \ref{apA:kernel-eq}.
		\begin{theorem}\label{apA:mercer-theorem}
			Let $\mu$ be a positive, separable Borel measure on a compact Hausdorff space $\X$ such that $\mu(\X) < \infty$ and $\mu(B) > 0$ for any open $B \subset \X$, $B \neq \varnothing$. If $\hat{A} \in \boundo$ is positive (i.e. $\langle u | \hat{A} u \rangle \geqslant 0$ for all $u \in \Hi = L_2(\X,\mu)$) and the associated kernel $K:\X \times \X \rightarrow \Comp$ is continuous, then
			\begin{equation}
			K(x,y) = \sum_{\lambda \in \sigma(\hat{A})} \sum_{i = 1}^{m_\lambda} \lambda u_{\lambda, i}(x) u_{\lambda, i}^*(y) ,
			\end{equation}
			where the convergence of the series on $\X \times \X$ is in the superior norm (i.e. $\| f \|_\infty := \sup_{x \in \X} \| f(x)\|$), $\lambda$ is an eigenvalue of $\hat{A}$, $m_{\lambda}$ is the multiplicity of the eigenvalue $\lambda$ and $\{u_{\lambda, i}\}^{i = 1,\cdots, m_{\lambda}}_{\lambda \in \sigma(\hat{A}) }$ is a basis of eigenvectors of $\hat{A}$.
		\end{theorem}
		Note that this theorem does not hold if $\X = \Rea^d$ and $\mu(dx) = dx$, since $\mu(\Rea^d) = \int_{\Rea^d}dx = \infty$. For this reason when $\X = \Rea^d$, whe shall often consider operators acting on $L_2(\Lambda)$, with $\Lambda \subset \Rea^d$ a compact set. More formally, defining the projector $\hat{P}_{\Lambda} : L_2(\Rea^d) \rightarrow L_2(\Lambda)$, given an operator $\hat{A} \in \mathcal{B}_2(L_2(\Rea^d))$ we can always consider its local version
		\begin{equation*}
		\hat{A}_{\Lambda} := \hat{P}_{\Lambda} \hat{A} \hat{P}_{\Lambda},
		\end{equation*}
		for which the Mercer theorem can be always applied. Note that in this case $K_{\Lambda}(x,y) = \chi_\Lambda(x)K(x,y) \chi_\Lambda(y)$ is the kernel of $\hat{A}_{\Lambda}$. Among all the Hilbert-Schmidt operators, an important subclass is the following.
		\begin{definition}
			Let $\Hi$ be a separable (possibly infinite-dimensional) Hilbert space. An operator $\hat{A} \in \boundo$ is called \emph{trace-class operator} if it satisfies one of the following equivalent conditions
			\begin{enumerate}
				\item[i)] There exist a basis $\{u_i\}_{i \in I}$ of $\Hi$ such that
				\begin{equation*}
				\sum_{i \in I} \langle u_i | \hat{A} u_i \rangle < \infty;
				\end{equation*}
				\item[ii)] $\sqrt{|\hat{A}|} \in \hso$;
				\item[iii)] $\hat{A} \in \compo$ and $\sum_{\lambda \in \sigma(\hat{A})} \lambda < \infty$.
			\end{enumerate}
		\end{definition}
		Note that here the Hilbert space must be separable (cfr. with Hilbert-Schmidt operator case). It can be proved that for $\Hi$ separable $\tco \subset \hso \subset \compo \subset \boundo$. As done for Hilbert-Schmidt operators, we can define a norm
		\begin{equation*}
		\| \hat{A} \|_1 := \bigg[\bigg\| \sqrt{|\hat{A}|} \bigg\|_2 \bigg]^2 = \sum_{\lambda \in \sigma(\hat{A})} \lambda,
		\end{equation*}
		which makes $\tco$ a Banach space, but this time it is not an Hilbert space (there is no inner product on $\tco$ inducing $\| \cdot \|_1$). We can see that $\| \cdot \|_1$ generalise the notion of trace to operators. In particular, the \emph{trace} operation
		\begin{equation*}
		Tr(\hat{A}) := \sum_{i \in I} \langle u_i | \hat{A} u_i \rangle,
		\end{equation*}
		where $\{u_i\}_{i \in I}$ is a basis of $\Hi$, is a well defined operation for every $\hat{A} \in \tco$. In addition one can prove that $Tr(\hat{A})$ does not depend on the basis chosen to compute it. Since $\tco \subset \hso$, we can use the Mercer theorem to compute the trace also in this case. In particular we have the following theorem \cite{brislawn1988kernels}.
		\begin{theorem}
			Let $K(x,y) \in L_2(\X \times \X, \mu \otimes \mu)$ be a continuous trace-class kernel on a $\sigma$-compact, locally compact space $\X$ and $\mu$ be a Radom measure. Then if $\hat{A}$ is the corresponding operator acting on $L_2(\X,\mu)$ we have
			\begin{equation*}
			Tr(\hat{A}) = \int_{\X} K(x,x) \mu(dx).
			\end{equation*}
		\end{theorem}
		The most common case is when $\X = \Rea^d$ (which is $\sigma$-compact and locally compact) and $\mu(dx) = dx$ (which is a Radom measure), then if $\hat{A}$ is of trace-class, we have
		\begin{equation*}
		Tr(\hat{A}) = \int_{\Rea^d} K(x,x)dx.
		\end{equation*}
		This theorem can be extended even to the case of non-continuous kernels, if they satisfy suitable conditions \cite{brislawn1991traceable}.
		\begin{definition}
			Let $\Hi = L_2(\X,\mu)$ be a separable Hilbert space and consider an operator $\hat{A}: \Hi \rightarrow \Hi$. If for any compact $\Lambda \subset \X$, $Tr(\hat{A}_\Lambda) < \infty$ we say that $\hat{A}$ is a \emph{locally trace-class operator}.
		\end{definition}
		The set of all locally trace-class operators on $\Hi$ will be labeled by $\mathcal{B}_1^{loc}(\Hi)$. It is not difficult to see that $\tco \subset \mathcal{B}_1^{loc}(\Hi)$ and that in general $\mathcal{B}_1^{loc}(\Hi) \not\subset \hso$, since the existence of the trace of an operator on a compact subset of $\X$ implies that the operator is Hilbert-Schmidt only locally. If $\X = \Rea^d$ and $\hat{A} \in \mathcal{B}_1^{loc}(L_2(\Rea^d))$, then the Mercer theorem always applies and the kernel of $\hat{A}$ on $\Lambda \subset \Rea^d$ is
		\begin{equation*}
		K_{\Lambda}(x,y) = \sum_{\mu_\Lambda} \mu_\Lambda u_{\mu_\Lambda}(x) u_{\mu_\Lambda}^*(y),
		\end{equation*}
		where $\mu_\Lambda$ and $u_{\mu_\Lambda}(x)$ are eigenvalues and eigenvectors of $\hat{A}_{\Lambda} = \hat{P}_{\Lambda} \hat{A} \hat{P}_{\Lambda}$ and the sum must be understood over the eigenvalues and their multiplicities (as in the statement of the Mercer theorem). However, when a locally trace-class operator is also an Hilbert-Schmidt operator (globally not only when restricted to some compact $\Lambda \subset \X$) the following theorem holds \cite{decreusefond2016determinantal}.
		\begin{theorem}\label{apA - LocallyTC-theorem}
			Let $\hat{A} \in \mathcal{B}_1^{loc}(L_2(\X,\mu)) \cap \mathcal{B}_2(L_2(\X,\mu))$ be a non negative operator, i.e. $\hat{A} > \hat{\mathbb{O}}$. Then one can choose its kernel $K(x,y)$ (defined everywhere, not only on $\Lambda \subset \X$ compact) such that
			\begin{enumerate}
				\item[i)] $K(x,y)$ is non-negative, i.e. $\sum_{i,j = 1}^n z_i^*  K(x_i, x_j) z_j \geqslant 0$, $\mu$-a.s., for any $x_1,\cdots, x_n \in \X$ and any $z_1,\cdots, z_n \in \Comp$;
				\item[ii)] $y\mapsto K_x(y):=K(y, x) \in L_2(\X,\mu)$, $\mu$-a.s.,  for any $x \in \X$ fixed;
				\item[iii)] For any compact $\Lambda \subset \X$, then
				\begin{equation*}
				Tr(\hat{P}_{\Lambda}\hat{A}\hat{P}_{\Lambda}) = \int_{\Lambda} K(x,x) \mu(dx)
				\end{equation*}
				and
				\begin{equation*}
				Tr(\hat{P}_{\Lambda} \hat{A}^n \hat{P}_{\Lambda}) = \int_{\Lambda} \langle K_x, \hat{A}^{n-2} K_x \rangle_{L_2(\Lambda,\mu)} \mu(dx) 
				\end{equation*}
				for $k \geqslant 2$.
			\end{enumerate}
		\end{theorem}
		
		The notion of trace-class operator allows to define the so called \emph{Fredholm’s determinant}, namely the determinant of an operator which differs from the identity by a trace-class operator. Before giving the general definition for this object, let us explain the motivation of this definition. Let $\Hi$ be a finite dimensional Hilbert space with $\dim \Hi = N < \infty$ and consider the tensor product Hilbert space 
		\begin{equation*}
		\Hi^{\otimes m} := \mbox{span}_{k_1,\cdots, k_m}\{ u_{k_1}^1 \otimes \cdots \otimes u_{k_m}^m  \}
		\end{equation*}
		where $\Hi = \mbox{span}_{k_j}\{u_{k_j}^j\}$ for all $j = 1,\cdots,m$. On  the tensor product Hilbert space, the scalar product is defined as
		\begin{equation*}
		\langle \Psi | \Phi \rangle_{\Hi^{\otimes m}} = \prod_{i =1}^m \langle \psi_i | \phi_i \rangle_{\Hi}
		\end{equation*}
		where $\Psi = \psi_1 \otimes \cdots \otimes \psi_m ,\Phi = \phi_1 \otimes \cdots \otimes \phi_m \in \Hi^{\otimes m}$. If $\hat{A}$ is an operator acting on $\Hi$, it can be extended to $\Hi^{\otimes m}$ by simply setting
		\begin{equation*}
		\hat{A}^{\otimes m} : \psi_1\otimes \cdots \otimes \psi_m \mapsto \hat{A}\psi_1 \otimes \cdots \otimes \hat{A}\psi_m
		\end{equation*}
		A natural subspace of $\Hi^{\otimes m}$ is the Hilbert space of the \emph{alternating power (exterior) Hilbert space} $\Hi^{\wedge m}$, which is the subspace of $\Hi^{\otimes m}$ of all the elements $\psi_1 \wedge \cdots \wedge \psi_m$ defined as
		\begin{equation*}
		\psi_1 \wedge \cdots \wedge \psi_m := \frac{1}{\sqrt{m!}} \sum_{\sigma \in P_m} \mbox{sign}(\sigma) \psi_{\sigma(1)} \otimes \cdots \otimes \psi_{\sigma(m)}.
		\end{equation*}
		The inner product on $\Hi^{\otimes m}$ can be used to induce an inner product on $\Hi^{\wedge m}$, and it turns out to be
		\begin{equation*}
		\langle \psi_1 \wedge \cdots \wedge \psi_m | \phi_1 \wedge \cdots \wedge \phi_m \rangle_{\Hi^{\wedge m}} = \det( [\langle \psi_i | \phi_j \rangle ]_{i,j = 1,\cdots, m} ) .
		\end{equation*}
		The action of an operator $\hat{A}$ can be extended on $\Hi^{\wedge m}$ by simply restricting $\hat{A}^{\otimes m}$ on $\Hi^{\wedge m}$, namely
		\begin{equation*}
		\hat{A}^{\wedge m}: \psi_1\otimes \cdots \otimes \psi_m \mapsto \hat{A}\psi_1 \wedge \cdots \wedge \hat{A}\psi_m.
		\end{equation*}
		Let $\hat{A} \in \tco$, then it is not difficult to see that $\hat{A}^{\wedge m} \in \mathcal{B}_1(\Hi^{\wedge m})$, thus the trace is well defined. After some combinatorics, one obtains
		\begin{equation*}
		Tr(\hat{A}^{\wedge m}) = \sum_{i_1 < \cdots < i_m} \lambda_{i_1} \cdots \lambda_{i_m},
		\end{equation*}
		where $\lambda_{i_k}$ the $i$-th eigenvalue of $\hat{A}$ in the $k$-th position of $\hat{A}\psi_1 \wedge \cdots \wedge \hat{A}\psi_m$. Now, summing over $m$, one obtains the following relation
		\begin{equation*}
		\begin{split}
		\sum_{m = 0}^N Tr(\hat{A}^{\wedge m}) &= \sum_{m = 0}^N \sum_{i_1 < \cdots < i_m} \lambda_{i_1} \cdots \lambda_{i_m} \\
		&= \prod_{i =1}^N (1 + \lambda_i).
		\end{split}
		\end{equation*}
		On the other hand, since the determinant of a matrix does not depend on the basis, we have that
		\begin{equation*}
		\det (\Id + \hat{A}) = \prod_{i = 1}^N (1 + \lambda_i),
		\end{equation*}
		which suggests that we can define the determinant of $\Id + \hat{A}$ as
		\begin{equation*}
		\det(\Id + \hat{A}) = \sum_{m = 0}^{\dim \Hi} Tr(\hat{A}^{\wedge m}).
		\end{equation*}
		The discussion done till here should justify the following definition.
		\begin{definition}
			Let $\hat{A} \in \tco$, then
			\begin{equation*}
			\det(\Id + \hat{A}) := \sum_{m = 0}^{\dim \Hi} Tr(\hat{A}^{\wedge m}).
			\end{equation*}
			where $\Hi$ is a (possibly infinite-dimensional) Hilbert space.
		\end{definition}
		The determinant defined above is called \emph{Fredholm’s determinant of $\Id + \hat{A}$}. An equivalent formula is the following. Given $\hat{A} \in \tco$ and $z \in \Comp$, then
		\begin{equation*}
		\det(\Id + z \hat{A}) = \exp \bigg( \sum_{m = 0}^{\dim \Hi} \frac{(-1)^{m-1} z^m Tr(\hat{A}^{\wedge m})}{m} \bigg)
		\end{equation*}
		which is known as \emph{Plemelj's formula}. Note that the above expression converges if $Tr(|\hat{A}|) < 1$. Finally, an important result is the following.
		\begin{theorem}
			If $\hat{A} \in \mathcal{B}_1(L_2(\Rea^d))$ and $K(x,y)$ is the associated kernel, then if $K(x,y): \Rea^d \times \Rea^d \rightarrow \Comp$ is continuous on $\Rea^d \times \Rea^d$, we have
			\begin{equation*}
			\begin{split}
			Tr( &\hat{A}^{\wedge m}) = \\
			&\frac{1}{m!} \int_{\Rea^d} \cdots \int_{\Rea^d} \det( [K(\xi_i,\xi_j)]_{i,j = 1, \cdots, m} )d\xi_1 \cdots d\xi_m.
			\end{split}
			\end{equation*}
			Thus
			\begin{equation*}
			\begin{split}
			\det(&\Id + \hat{A}) =  1 + \\
			&\sum_{m =0}^\infty \frac{1}{m!} \int_{\Rea^d} \cdots \int_{\Rea^d} \det( [K(\xi_i,\xi_j)]_{i,j = 1, \cdots, m} )d\xi_1 \cdots d\xi_m.
			\end{split}
			\end{equation*}
		\end{theorem}
		In this last theorem, also called \emph{Fredholm’s series expansion}, the writing $[K(\xi_i,\xi_j)]_{i,j = 1, \cdots, m}$ is a short hand notation for the $m \times m$ matrix whose $(i,j)$-th entry is $K(\xi_i,\xi_j)$. 
		
		\subsection*{B. Dirichlet forms and Markov processes }
		
		Let $\Hi$ be an Hilbert space and consider a dense subspace $\mathcal{D} \subset \Hi$. A map $\varepsilon: \mathcal{D} \times \mathcal{D} \rightarrow \Rea$ is said \emph{real bilinear form} if $\varepsilon(\alpha u +\beta v, z) = \alpha \varepsilon(u,z) + \beta \varepsilon(v,z)$, for any $u,v,z \in \mathcal{D}$, and $\alpha,\beta \in \Rea$. A real bilinear form is said \emph{positive} if $\varepsilon(u,u) \geqslant 0$, for any $u \in \mathcal{D}$, and \emph{symmetric} if $\varepsilon(u,v) = \varepsilon(v,u)$ for any $u,v \in \mathcal{D}$. Let $\langle \cdot | \cdot \rangle$ denote the scalar product on $\Hi$, and consider the bilinear form $(u,v) := \varepsilon(u,v) + \langle u | v \rangle$ which is defined for all $u,v \in \mathcal{D}$. When $\varepsilon$ is symmetric and positive, the bilinear form $(\cdot,\cdot)$ is an inner product on $\mathcal{D}$. We say that the real symmetric positive bilinear form $\varepsilon$ is \emph{closed} if $\mathcal{D}$ is an Hilbert space, when equipped with the inner product $(\cdot,\cdot)$ defined above. Now we are ready to define a Dirichlet form.
		\begin{definition}
			Let $(\X,\mathcal{X},\mu)$ be a $\sigma$-finite measure space and set $\Hi = L_2(\X,\mu)$. Consider a real symmetric positive bilinear form $\varepsilon:\mathcal{D} \times \mathcal{D} \rightarrow \Rea$, where $\mathcal{D} \subset \Hi$ is dense, which is also closed. If for any $\epsilon > 0$, there exists a real function $\phi_{\epsilon}(x)$, $x \in \Rea$ with the following features
			\begin{enumerate}
				\item[i)] $\phi_{\epsilon}(x) = x$ for $x \in [0,1]$,
				\item[ii)] $\phi_{\epsilon}(x) \in [-\epsilon, 1+ \epsilon]$ for any $x \in \Rea$,
				\item[iii)] $\phi_{\epsilon}(x) - \phi_{\epsilon}(x') \in [0,x-x']$ whenever $x' < x$,
			\end{enumerate}
			for which when $u \in \mathcal{D}$ then $\phi_{\epsilon}(x) \in \mathcal{D}$ and
			\begin{equation*}
			\varepsilon(\phi_{\varepsilon}(u),\phi_{\varepsilon}(u)) \leqslant \varepsilon(u,u),
			\end{equation*}
			the couple $(\varepsilon,\mathcal{D})$ is said \emph{Dirichlet form} on $L_2(\X,\mu)$.
		\end{definition}
		For the rest of this appendix, we assume that $\X$ is a locally compact second countable metric space and $\mu$ is a Borel measure having support on the whole $\X$.
		This abstract object is important because of the following theorem (Th 1.3.1, \cite{fukushima2010dirichlet})
		\begin{theorem}
			There is a one to one correspondence between a Dirichlet form $(\varepsilon,\mathcal{D})$ on $L_2(\X,\mu)$ and the family of non-positive definite self-adjoint operators on $L_2(\X,\mu)$. The correspondence is the following
			\begin{equation*}
			\varepsilon(u,v) = \langle [-\hat{H}]^{1/2} u | [-\hat{H}]^{1/2} v \rangle
			\end{equation*}
			and $D([-\hat{H}]^{1/2}) = \mathcal{D}$, where $D(\hat{H})$ is the domain of the operator $\hat{H}$.
		\end{theorem}
		The operator $\hat{H}$ is called \emph{generator} of the Dirichlet form. Note that, because it is self-adjoint we can write
		\begin{equation*}
		\varepsilon(u,v) = -\langle u |\hat{H} v \rangle
		\end{equation*}
		but note that $\mbox{Dom}(\hat{H}) \subset \mathcal{D}$. Using such a generator, one can define the operator
		\begin{equation}\label{DF-generator}
		\hat{T}_t := \exp( - t \hat{H})
		\end{equation}
		acting on $L_2(\X,\mu)$. Since $-\hat{H}$ is non-negative, $\hat{T}_t$ is always bounded. It also has the \emph{semigroup property}, i.e. $\hat{T}_{t+s} = \hat{T}_t \hat{T}_s$, and it can be proved that it is also \emph{strongly continuous} on $L_2(\X,\mu)$. The link between these objects and stochastic processes is encoded in the following theorem (Th. 1.4.1, \cite{fukushima2010dirichlet}).
		\begin{theorem}\label{DF - Th. gen}
			Let $\varepsilon$ be a Dirichelt form on $L_2(\X,\mu)$ with generator $\hat{H}$. Then the operator \eqref{DF-generator} is a strongly continuous semigroup such that
			\begin{equation*}
			0 \leqslant \hat{T}_t u \leqslant 1 \qquad\mu\mbox{-a.s}
			\end{equation*}
			whenever $0 \leqslant u \leqslant 1$, $\mu$-a.s., with $u \in L_2(\X,\mu)$.
		\end{theorem}
		At this point the connection with stochastic processes starts to appear. Given a Markov process, $\{X_t\}_{t \in \Rea^+}$ taking values on $\X$ with distribution $\mu_X$, consider its transition probability density $p(x,t|s,y)$. Note that we are implicitly assuming that the transition probability admits a density, however the whole argument remains valid even in the general case. Given $p(x,t|s,y)$, we can define the following integral operator
		\begin{equation*}
		\hat{S}_t f(x) := \int_{\X} f(y)p(x,t|s,y)dy,
		\end{equation*}
		where $f(x)$ is a bounded measurable function. More generally this operator is well defined any for $f \in L_2(\X,\mu_X)$. From the Markov property, the semigroup property follows, i.e. $\hat{S}_{t+s} = \hat{S}_t \hat{S}_s$. We note that $0 <\hat{S}_r f(x) < 1$ for all $x \in \X$ whenever $0 < f(x) < 1$ for all $x \in \X$, i.e. when $f(x)$ is a probability density. In addition $\hat{S}_t 1 =1$, which is a consequence of the fact that $p(x,t|s,y)$ are transition probability densities. Finally, one can prove that $\hat{S}_t$ is strongly continuous in $t$ when thought as a linear operator on $L_2(\X,\mu_X)$. Taking the generator of $\hat{S_t}$, i.e. the operator
		\begin{equation*}
		-\hat{H}' := \lim_{t \rightarrow 0} \frac{\hat{S}_t f(x) - f(x)}{t}
		\end{equation*}
		where $f(x) \in L_2(\X,\mu_X)$, one can define a Dirichlet form $\varepsilon'(f,g) = - \langle f | \hat{H}' g \rangle$ on $L_2(\X,\mu_X)$. Thus one can study the properties of the Markov process $\{X_t\}_{t \in \Rea^+}$ using Dirichlet forms. Note that the opposite is not always true: for example in theorem \ref{DF - Th. gen}, nothing is said on the condition $\hat{T}_t 1 =1$ which clearly holds for a Markov process. Among the properties of $\{X_t\}_{t \in \Rea^+}$ that can studied using Dirichlet forms, there are also the path properties. In particular one can verify if the process is a diffusion or not. We conclude by saying that to be sure to obtain a Markov process, one needs to add other conditions on the Dirichlet form $(\varepsilon,\mathcal{D})$: for example, to obtain $\hat{T}_t 1 = 1$, one has to require that $1 \in \mathcal{D}$ and $\varepsilon(1,1) =0$. For a detailed and complete discussion on the general relation between Dirichlet forms and Markov processes we refer to \cite{fukushima2010dirichlet}.

			\bibliographystyle{unsrt}
			\bibliography{bib-c.bib}

\begin{thebibliography}{10}

\bibitem{LC2}
Luca Curcuraci.
\newblock On non-commutativity in quantum theory (ii): toy models for
  non-commutative kinematics.
\newblock arXiv:1803.04916 [quant-ph], 2018.

\bibitem{baddeley2007spatial}
Adrian Baddeley, Imre B{\'a}r{\'a}ny, and Rolf Schneider.
\newblock Spatial point processes and their applications.
\newblock {\em Stochastic Geometry: Lectures given at the CIME Summer School
  held in Martina Franca, Italy, September 13--18, 2004}, pages 1--75, 2007.

\bibitem{daley2007introduction}
Daryl~J Daley and David Vere-Jones.
\newblock {\em An introduction to the theory of point processes}.
\newblock Springer Science \& Business Media, 2007.

\bibitem{macchi1975coincidence}
Odile Macchi.
\newblock The coincidence approach to stochastic point processes.
\newblock {\em Advances in Applied Probability}, 7(01):83--122, 1975.

\bibitem{hough2009zeros}
John~Ben Hough, Manjunath Krishnapur, Yuval Peres, and B{\'a}lint Vir{\'a}g.
\newblock {\em Zeros of Gaussian analytic functions and determinantal point
  processes}, volume~51.
\newblock American Mathematical Society Providence, RI, 2009.

\bibitem{hough2006determinantal}
J~Ben Hough, Manjunath Krishnapur, Yuval Peres, B{\'a}lint Vir{\'a}g, et~al.
\newblock Determinantal processes and independence.
\newblock {\em Probability surveys}, 3(206-229):9, 2006.

\bibitem{decreusefond2016determinantal}
Laurent Decreusefond, Ian Flint, Nicolas Privault, and Giovanni~Luca Torrisi.
\newblock Determinantal point processes.
\newblock In {\em Stochastic Analysis for Poisson Point Processes}, pages
  311--342. Springer, 2016.

\bibitem{soshnikov2000determinantal}
Alexander Soshnikov.
\newblock Determinantal random point fields.
\newblock {\em Russian Mathematical Surveys}, 55(5):923--975, 2000.

\bibitem{shirai2003random}
Tomoyuki Shirai and Yoichiro Takahashi.
\newblock Random point fields associated with certain fredholm determinants i:
  fermion, poisson and boson point processes.
\newblock {\em Journal of Functional Analysis}, 205(2):414--463, 2003.

\bibitem{Kui}
Arno~BJ Kuijlaars.
\newblock Lecture notes in: Riemann-hilbert problems and multiple orthogonal
  polynomials.
\newblock
  \url{https://math.vanderbilt.edu/dept/conf/constructive2014/KuijlaarsNotes.pdf},
  2014.

\bibitem{osada1996dirichlet}
Hirofumi Osada.
\newblock Dirichlet form approach to infinite-dimensional wiener processes with
  singular interactions.
\newblock {\em Communications in mathematical physics}, 176(1):117--131, 1996.

\bibitem{yoo2005dirichlet}
Hyun~Jae Yoo.
\newblock Dirichlet forms and diffusion processes for fermion random point
  fields.
\newblock {\em Journal of Functional Analysis}, 219(1):143--160, 2005.

\bibitem{decreusefond2012stochastic}
Laurent Decreusefond, Ian Flint, Nicolas Privault, and Giovanni~Luca Torrisi.
\newblock Stochastic dynamics of determinantal processes by integration by
  parts.
\newblock {\em arXiv preprint arXiv:1210.6109}, 2012.

\bibitem{LC}
Luca Curcuraci.
\newblock On non-commutativity in quantum theory (i): from classical to quantum
  probaility.
\newblock arXiv:1803.04913 [quant-ph], 2018.

\bibitem{friesecke2009ehrenfest}
Gero Friesecke and Mario Koppen.
\newblock On the ehrenfest theorem of quantum mechanics.
\newblock {\em Journal of Mathematical Physics}, 50(8):082102, 2009.

\bibitem{bargmann1954unitary}
Valentine Bargmann.
\newblock On unitary ray representations of continuous groups.
\newblock {\em Annals of Mathematics}, pages 1--46, 1954.

\bibitem{moretti2013spectral}
Valter Moretti.
\newblock {\em Spectral theory and quantum mechanics: with an introduction to
  the algebraic formulation}.
\newblock Springer Science \& Business Media, 2013.

\bibitem{aerts1978physical}
Diederik Aerts and Ingrid Daubechies.
\newblock Physical justification for using the tensor product to describe two
  quantum systems as one joint system.
\newblock {\em Helv. Phys. Acta}, 51(5-6), 1978.

\bibitem{bell1964einstein}
John~S Bell.
\newblock On the einstein podolsky rosen paradox, 1964.

\bibitem{pusey2012reality}
Matthew~F Pusey, Jonathan Barrett, and Terry Rudolph.
\newblock On the reality of the quantum state.
\newblock {\em Nature Physics}, 8(6):475--478, 2012.

\bibitem{khrennikov2009contextual}
Andrei~Y Khrennikov.
\newblock {\em Contextual approach to quantum formalism}, volume 160.
\newblock Springer Science \& Business Media, 2009.

\bibitem{khrennikov2016probability}
Andrei Khrennikov.
\newblock {\em Probability and Randomness: Quantum versus Classical}.
\newblock World Scientific, 2016.

\bibitem{khrennikov2014bell}
Andrei Khrennikov.
\newblock Bell as the copernicus of probability.
\newblock {\em arXiv preprint arXiv:1412.6987}, 2014.

\bibitem{frederick1976stochastic}
Carlton Frederick.
\newblock Stochastic space-time and quantum theory.
\newblock {\em Physical Review D}, 13(12):3183, 1976.

\bibitem{roy2012statistical}
Sisir Roy.
\newblock {\em Statistical geometry and applications to microphysics and
  cosmology}, volume~92.
\newblock Springer Science \& Business Media, 2012.

\bibitem{bombelli2009semiclassical}
Luca Bombelli, Alejandro Corichi, and Oliver Winkler.
\newblock Semiclassical quantum gravity: obtaining manifolds from graphs.
\newblock {\em Classical and Quantum Gravity}, 26(24):245012, 2009.

\bibitem{bombelli1987space}
Luca Bombelli, Joohan Lee, David Meyer, and Rafael~D Sorkin.
\newblock Space-time as a causal set.
\newblock {\em Physical review letters}, 59(5):521, 1987.

\bibitem{brightwell1991structure}
Graham Brightwell and Ruth Gregory.
\newblock Structure of random discrete spacetime.
\newblock {\em Physical review letters}, 66(3):260, 1991.

\bibitem{reid2003manifold}
David~D Reid.
\newblock Manifold dimension of a causal set: Tests in conformally flat
  spacetimes.
\newblock {\em Physical Review D}, 67(2):024034, 2003.

\bibitem{aerts2001probability}
S.~{Aerts}.
\newblock {Probability Conservation and the State Determination Problem}.
\newblock In A.~Khrennikov, editor, {\em Foundations of Probability and
  Physics}, pages 39--49, December 2001.

\bibitem{fukushima2010dirichlet}
Masatoshi Fukushima, Yoichi Oshima, and Masayoshi Takeda.
\newblock {\em Dirichlet forms and symmetric Markov processes}, volume~19.
\newblock Walter de Gruyter, 2010.

\bibitem{nagasawa2012schrodinger}
Masao Nagasawa.
\newblock {\em Schr{\"o}dinger equations and diffusion theory}, volume~86.
\newblock Birkh{\"a}user, 2012.

\bibitem{nagasawa2012stochastic}
Masao Nagasawa.
\newblock {\em Stochastic processes in quantum physics}, volume~94.
\newblock Birkh{\"a}user, 2012.

\bibitem{nelson1966derivation}
Edward Nelson.
\newblock Derivation of the schr{\"o}dinger equation from newtonian mechanics.
\newblock {\em Physical review}, 150(4):1079, 1966.

\bibitem{nagasawa1989transformations}
Masao Nagasawa.
\newblock Transformations of diffusion and schr{\"o}dinger processes.
\newblock {\em Probability Theory and Related Fields}, 82(1):109--136, 1989.

\bibitem{karambal2010introductory}
Issa Karambal, Veerle Ledoux, Simon~JA Malham, and Jitse Niesen.
\newblock Introductory fredholm theory and computation.
\newblock 2010.

\bibitem{brislawn1988kernels}
Chris Brislawn.
\newblock Kernels of trace class operators.
\newblock {\em Proceedings of the American Mathematical Society},
  104(4):1181--1190, 1988.

\bibitem{brislawn1991traceable}
Christopher Brislawn.
\newblock Traceable integral kernels on countably generated measure spaces.
\newblock {\em Pacific Journal of Mathematics}, 150(2):229--240, 1991.

\end{thebibliography}

	\end{document}